\documentclass[twocolumn,aps,prb,superscriptaddress,showpacs]{revtex4}

\usepackage{graphicx}
\usepackage{amsmath}
\usepackage{amsfonts}
\usepackage{amssymb}
\usepackage{latexsym}
\usepackage{dcolumn}

\begin{document}

\title{First-principles study of the dipole layer formation at metal-organic interfaces}

\author{Paul C. Rusu}
\affiliation{Computational Materials Science, Faculty of Science and Technology and MESA+ Institute
for Nanotechnology, University of Twente, P.O. Box 217, 7500 AE Enschede, The Netherlands.}
\author{Gianluca Giovannetti}
\affiliation{Consiglio Nazionale delle Ricerche - Istituto Nazionale per la Fisica della Materia (CNR-INFM), CASTI Regional Laboratory, 67100 L'Aquila, Italy.}
\author{Christ Weijtens}
\affiliation{Philips Research Laboratories Aachen, Weisshausstrasse 2, D-52066 Aachen, Germany.}
\author{Reinder Coehoorn}
\affiliation{Philips Research Laboratories, Prof. Holstlaan 4, 5656 AA Eindhoven, The Netherlands.}
\author{Geert Brocks}
\affiliation{Computational Materials Science, Faculty of Science and Technology and MESA+ Institute
for Nanotechnology, University of Twente, P.O. Box 217, 7500 AE Enschede, The Netherlands.}
\date{\today}


\begin{abstract}
We study the dipole layer formed at metal-organic interfaces by means of first-principles
calculations. Interface dipoles are monitored by calculating the work function change of Au, Ag,
Al, Mg and Ca surfaces upon adsorption of a monolayer of PTCDA
(3,4,9,10-perylene-tetra-carboxylic-di-anhydride), perylene or benzene molecules. Adsorption of
PTCDA leads to pinning of the work function for a range of metal substrates. It gives interface
dipoles that compensate for the difference in the clean metal work functions, leading to a nearly
constant work function. In contrast, adsorption of benzene always results in a decrease of the
work function, which is relatively constant for all metal substrates. Both effects are found in
perylene, where adsorption on low work function metals gives work function pinning, whereas
adsorption on high work function metals gives work function lowering. The work function changes
upon adsorption are analyzed and interpreted in terms of two competing effects. If the molecule
and substrate interact weakly, the molecule pushes electrons into the surface, which lowers the
work function. If the metal work function is sufficiently low with respect to the unoccupied
states of the molecule, electrons are donated into these states, which increases the binding and
the work function.
\end{abstract}

\maketitle

\section{Introduction}

Applications of organic semiconductors in light-emitting
diodes,\cite{Tang:appl_lett,Burroughes:nature} field-effect
transistors\cite{Tsumura:appl_lett,Garnier:science} and solar
cells\cite{Sariciftci:science,Yu:adv_mat} have stimulated research into the fundamental electronic
properties of organic materials and their interfaces with metal electrodes.\cite{Gershenson:rmp06,Kahn:jpsb03} The weak forces between the molecules in an organic material
lead to small band widths, which enhances the importance of electron-phonon and electron-electron
interactions.\cite{Brocks:sym99,Brocks:prl04,Giovannetti:prb08,Craciun:prb09} Nevertheless high charge carrier mobilities can be
achieved in well-ordered molecular crystals.\cite{Gershenson:rmp06} As the quality of molecular
crystals increases, transport of charge carriers across the interfaces between metal electrodes
and the organic material starts to determine the performance of the devices.\cite{Hulea:apl06}
Metal organic interfaces (MOIs) often give rise to a non-Ohmic behavior, indicating the existence
of significant Schottky barriers.

Chemical bonding between molecules and metal surfaces modifies the charge distribution at a MOI. It
results in an interface dipole layer, which strongly influences the Schottky barrier height.\cite{Tung:prl00,Tung:prb01,Rusu:prb06} This effect of chemical bonding is observed very clearly
in self-assembled monolayers of thiolate molecules chemically bonded to noble metal surfaces.\cite{Rusu:prb06,Derenzi:prl05,Bredas:prl06,Rusu:jpcb06,Bredas:nanolett07,Rusu:jpcc07} Common
organic semiconductors however consist of closed shell molecules, which are usually thought to bind
weakly to metal surfaces. It has therefore been assumed for a long time that the charge reordering
at such MOIs is insignificant and that no appreciable interface dipole is formed.

In absence of an interface dipole, the Schottky barrier at a MOI can be predicted by aligning the
vacuum levels of the metal and the organic material, called the Schottky-Mott rule. Over the last
decade, however, experimental studies have indicated the general breakdown of the vacuum alignment
rule, and have demonstrated that significant interface dipoles are formed at MOIs.
\cite{Kahn:jpsb03,Hill:apl98,Ishii:adv_mat,Koch:prl05,Knupfer:ass05,Ishii:pssa04} In addition such
studies have shown that interface dipoles at MOIs are localized foremost at the first molecular
layer covering the metal. The interface dipoles are not affected much by deposition of additional
organic layers. Because MOI dipoles are localized at the interface, they can be extracted from the
change in the surface work function after deposition of a single organic layer.

Ideas inspired by chemical bonding have been put forward to explain large
interface dipoles. Conventional semiconductors such as Si have reactive surfaces, which bind
strongly to metal overlayers. A significant density of states is then often created at the
metal-semiconductor interface within the band gap of the semiconductor, the so-called metal induced
gap states (MIGS).\cite{Heine:phys_rev,Louie:prb,Flores:jpc77} In this model, MIGS determine the
charge distribution at the interface and hence the interface dipole. The MIGS model has also been
applied to MOIs.\cite{Vazquez:epl04,Vazquez:ass04,Vazquez:jcp07} It requires a strong interaction
between the metal and the organic material.

If molecules are physisorbed onto a metal surface, one expects a relatively weak interaction between the
molecular semiconductor and the metal. For physisorbed molecules interface dipoles at MOIs have
been explained by the so-called pillow effect.\cite{Silva:prl,Bagus:prl02,Crispin:jacs,Morikawa:prb04,Bagus:apl05} If a molecule approaches a
metal surface, the electronic clouds of the molecule and the metal start to overlap. The Pauli
exchange repulsion between these clouds leads to a spatial redistribution of electrons, which
modifies the surface dipole. Since the electronic cloud of the metal is usually ``softer'' than that
of the molecule, the net effect of Pauli repulsion is that electrons are pushed back into the
metal. The result is an interface dipole that decreases the work function of the surface.

A decrease of the work function is commonly found if inert atoms or small molecules are adsorbed on a metal surface.\cite{Silva:prl,Bagus:prl02,Crispin:jacs,Morikawa:prb04,Bagus:apl05,Chen:prb84} Remarkably, adsorption of larger, $\pi$-conjugated, molecules can lead to a substantial increase, as well as a decrease of the work function and the dependence of this work function shift and the associated interface dipole on the molecules and the metal has
been the subject of intensive experimental study.\cite{Kahn:jpsb03,Hill:apl98,Ishii:adv_mat,Koch:prl05,Knupfer:ass05,Ishii:pssa04} If the work function $W$ of a surface after coverage with a
molecular layer is measured for a range of metal substrates with different initial work functions
$W_c$, the results can be characterized by the parameter
\begin{equation}
S = \frac{dW}{dW_c}, \label{S_parameter}
\end{equation}
where $W_c$, $W$ are the work functions of the clean metal surface and of the surface with the
adsorbed organic layer, respectively.

The vacuum level alignment, or Schottky-Mott rule gives $S=1$. Assuming that the pillow
effect does not depend strongly on the metal substrate, it gives a relatively
constant decrease of the work function, leading to $S\approx 1$. Although this is observed for some
molecules, very often $S$ is significantly smaller than 1.\cite{Kahn:jpsb03} Moreover, there is no
a priori reason why $S$ should be a constant. Indeed for some molecules and polymers several regimes can be
distinguished, between which a transition from $S\approx 1$ to $S\approx 0$ is observed.\cite{Tang:cpl04,Tengstedt:apl06} For the case where the organic layer is separated from the metal electrode by a thin insulating barrier, this behavior is interpreted with a model that assumes charge transfer across this barrier.\cite{Fahlman:jpcm07,Braun:advmat09,Davids:jap95}

In this paper we study the dipole formation at interfaces of monolayers of PTCDA
(3,4,9,10-perylene-tetra-carboxylic-di-anhydride), perylene and benzene molecules adsorbed on
close-packed metal surfaces of Au, Ag, Al, Mg and Ca by means of density functional theory (DFT)
calculations. We have selected these surfaces because they have a similar and simple structure, as well as a simple electronic (free electron like) structure. Yet their work functions span a range from 3.0 eV (Ca) to 5.3 eV (Au), allowing to study the effect of the metal work function on the interface dipole. A preliminary account of this work has been given in Ref.~\onlinecite{Rusu:jpcc09}.

The molecules are chosen on account of their difference in complexity and their different behavior experimentally. PTCDA is a fairly complex conjugated molecule with a relatively small electronic gap. From an experimental point of view a PTCDA monolayer on metal surfaces has been a model system to study MOIs. Deposition of PTCDA on noble metal surfaces leads to well-ordered epitaxial overlayers.\cite{Forrest:chemrev97} In particular the structure and electronic structure of PTCDA on Ag(111) have been studied intensively.\cite{Umbach:surf_sci,Eremtchenko:nat03,Eremtchenko:njp04,Hauschild:prl05,Zou:ss06,Kraft:prb06,Temirov:nat06,Gerlach:prb07,Romaner:njp09}
Work function measurements have been performed for PTCDA adsorbed on a range of metal surfaces.\cite{Kahn:jpsb03,Hill:apl98,Duhm:orgel08,Kawabe:orgel08} These measurements give a work function that is roughly independent of the metal substrate, i.e. $S\approx 0$, meaning that adsorption of a PTCDA monolayer on a high work function metal gives a decrease of the work function, whereas adsorption on a low work function metal gives an increase.

In contrast, experimental data suggest that adsorption of the simple conjugated molecule benzene on a metal surface always gives a decrease of the work function, with $S$ in the range 0.6-1.0.\cite{Dudde:ss90,Zhou:ss90,Velic:jcp98,Duschek:cpl00,Gaffney:cp00,Bagus:apl05} The size and complexity of the perylene molecule is between that of benzene and PTCDA. The structure of an adsorbed perylene monolayer is thought to be similar to that of a PTCDA layer.\cite{Eremtchenko:njp04,Seidel:prb01,Eremtchenko:jmatres04} Depending on the metal substrate, the work function can decrease or increase, but there is no uniform pinning as for PTCDA.\cite{Yan:apl02}

This paper is organized as follows. In the next section we give the technical details of our
calculations. In Sec.~\ref{PTCDA} we present our results obtained for adsorption of PTCDA
monolayers on different metal surfaces. We compare results obtained with different density functionals, and study the influence of the packing density of the molecules on the surface.
Section~\ref{benzene_and_perylene} gives the results obtained for adsorbed benzene and perylene
monolayers. The results are discussed in Sec.~\ref{discussion} with the help of a simple phenomenological model, and a short summary and conclusions are given in Sec.~\ref{conclusions}.

\section{Computational details}\label{compdetails}

The electronic structure is treated within density functional theory (DFT)
\cite{Hohenberg:pr64,Kohn:pr65} using the local density approximation (LDA),\cite{Perdew:prb81,Ceperley:prl80} or the generalized gradient approximation (GGA) with the PW91
exchange-correlation functional.\cite{Perdew:prb92} The calculations are performed with the VASP
(Vienna \textit{ab initio} simulation package) program,\cite{Kresse:prb93, Kresse:prb96} which
uses the projector augmented wave (PAW) method.\cite{Kresse:prb99, Bloechl:prb94} For Au and Ag
atoms the outer shell $d$ and $s$ electrons are treated as valence electrons, for Al the outer
shell $s$ and $p$ electrons, and for Mg and Ca the outer shell $s$ electrons. For atoms of first
row elements the $2s$ and $2p$ electrons are treated as valence electrons. The valence pseudo wave
functions are expanded in a basis set consisting of plane waves. All plane waves up to a kinetic
energy cutoff of 400 eV are included.

To model the metal-molecule interface, we use a supercell containing a slab of at least three metal
layers with one layer of molecules adsorbed on one surface, and a vacuum region of at least 10 \AA.
Periodic boundary conditions are applied and the atomic positions in the top metal layer and in the
molecules are allowed to relax. A dipole correction is applied to avoid spurious interactions
between dipoles of repeated slabs along the direction normal to the surface.\cite{Neugebauer:prb92}

The electronic structure is calculated self-consistently using a $3\times 3$ (for  PTCDA) to $5\times 5$ (for perylene and benzene) $\mathbf{k}$-point grid in the
irreducible surface Brillouin zone (SBZ) according to the Monkhorst-Pack scheme
\cite{Monkhorst:prb76} and applying a Methfessel-Paxton smearing of 0.2 eV.\cite{Methfessel:prb89}
A $3\times 3$ $\mathbf{k}$-point grid gives well-converged results for PTCDA layers, because of the large size of
the surface unit cell (see next section). For accurate calculations of total energies and densities
of state (DOS) the charge densities are recalculated with an up $7\times 7$ $\mathbf{k}$-point grid,
using the tetrahedron method.\cite{tetrahedron:93} DOSs are plotted using Gaussian smearing with a
broadening parameter of 0.1 eV.

Work functions are evaluated from the expression
\begin{equation}
W = V(\infty)-E_{F}, \label{eq:w}
\end{equation}
where $V(\infty)$ is the electrostatic potential in the vacuum region and $E_{F}$ is the Fermi
energy of the bulk metal. $V(\infty)$ is obtained from the potential averaged in the $(x,y)$ plane
\begin{equation}
\overline{V}(z) = \frac{1}{A}\iint_{cell}V(x,y,z)dxdy , \label{eq:vav}
\end{equation}
where $V(x,y,z)$ is the electrostatic potential on a real space grid in the supercell. In practice
$\overline{V}(z)$ reaches an asymptotic value $V(\infty)$ at a distance of a few \AA\ from the
surface.\cite{Rusu:jpcb06,Rusu:jpcc07} An accurate value of $E_{F}$ is obtained from a separate
bulk calculation, following the procedure described in Ref.~\onlinecite{Fall:jp99}.

Test calculations regarding slab thickness, vacuum thickness, $\mathbf{k}$-point sampling grid and
plane waves kinetic energy cutoff are performed in order to estimate the convergence. From these
tests we find that with the parameters given above, total energies are converged to within 0.01 eV
and work functions to within 0.05 eV. The results for PTCDA on Ca(111) turn out to be the most
sensitive with respect to vacuum thickness and $\mathbf{k}$-point sampling. So for this system the
results have been obtained using a $5\times 5$ $\mathbf{k}$-point grid in the irreducible SBZ and a vacuum
thickness of at least 14 \AA.

To analyze our results we also use properties of isolated molecules, such as the electron affinity (EA), as will be
discussed in Sec.~\ref{discussion}. For calculations on
isolated molecules we use the GAMESS program,\cite{Schmidt:jcc93} and treat the electronic
structure within DFT using the BLYP functional.\cite{Becke:pra88,LYP:prb88} We use the 6-31+G$^*$ basis set, which gives EAs for acenes that are converged on a scale of $\sim 0.1$ eV.\cite{Modelli:jpca06} As in Ref.~\onlinecite{Modelli:jpca06}, we find that including a diffuse
orbital in the basis set is important and that the smaller 6-31G$^*$ basis set does not give a
sufficiently converged EA.\cite{Andrzejak:theochem00} For instance, the EAs of PTCDA obtained using the 6-31G$^*$ and 6-31+G$^*$ bases differ by 0.4 eV. The Kohn-Sham energy levels of the (neutral) isolated molecule calculated with VASP and GAMESS are very similar, illustrating that, provided the basis sets are sufficiently converged, the PW91 and BLYP functionals give similar results.

\section{PTCDA}\label{PTCDA}

Before discussing the results obtained for adsorbed layers, we benchmark our calculations on clean
metal surfaces. We consider the close-packed (111) surfaces of fcc Au, Ag, Al, and Ca, and the
(0001) surface of hcp Mg. The metals in this set are relatively simple, free electron like and the
set of surfaces spans a considerable range in work functions. Table~\ref{lattice constants} lists
the optimized nearest neighbor distances of the bulk metals, calculated with GGA(PW91) and LDA
functionals. As usual, the GGA values are larger than the LDA values, but both are generally in
reasonable agreement with experiment. We use these optimized distances to construct the surface
unit cells.

\begin{table}[btp]
\caption{Optimized nearest neighbor distances in the bulk metals. All values are in \AA.}
\label{lattice constants}
\begin{ruledtabular}
\begin{tabular}{cccccc}
     & Au & Ag & Al & Mg & Ca \\
\hline
GGA      &2.94   &2.93   &2.86   &3.19   &3.92\\
LDA      &2.87   &2.84   &2.82   &3.13   &3.78\\
exp.     &2.88   &2.89   &2.86   &3.21   &3.95\\
\end{tabular}
\end{ruledtabular}
\end{table}

\begin{table}[btp]
\caption{Calculated work functions of clean (111) surfaces; (0001) for Mg . All values are in eV.}
\label{clean_metal_wfs}
\begin{ruledtabular}
\begin{tabular}{cccccc}
     & Au & Ag & Al & Mg & Ca \\
\hline
GGA    &5.25   &4.50   &4.08   &3.74   &2.98\\
LDA    &5.52   &4.90   &4.21   &3.93   &3.08\\
exp.   &5.26,\footnotemark[1] 5.35\footnotemark[2]
       &4.46,\footnotemark[3] 4.50,\footnotemark[4] 4.56\footnotemark[5]
       &4.24\footnotemark[6]
       &3.78\footnotemark[7]
       &(2.87)\footnotemark[8]    \\
calc.  &5.27,\footnotemark[9] 5.35\footnotemark[2] & 4.42\footnotemark[10]
       &4.25\footnotemark[11] &3.76,\footnotemark[12] 3.88\footnotemark[13]
       &2.86\footnotemark[14] \\
\end{tabular}
\end{ruledtabular}
\footnotetext[1]{Ref.~\onlinecite{Hansson:prb78} }
\footnotetext[2]{Ref.~\onlinecite{Derenzi:prl05}}
\footnotetext[3]{Ref.~\onlinecite{Chelvayohan:jpc82}}
\footnotetext[4]{Ref.~\onlinecite{Monreal:cm03}    }
\footnotetext[5]{Ref.~\onlinecite{Giesen:prb87} } \footnotetext[6]{Ref.~\onlinecite{Grepstad:ss76}
} \footnotetext[7]{Ref.~\onlinecite{Anderson:prev} } \footnotetext[8]{polycrystalline value,
Ref.~\onlinecite{Michaelson:jap77}} \footnotetext[9]{GGA, Ref.~\onlinecite{Piccinin:jcp03} }
\footnotetext[10]{GGA, Ref.~\onlinecite{Bocquet:mol_phys} } \footnotetext[11]{LDA,
Ref.~\onlinecite{Curioni:IBM} } \footnotetext[12]{LDA, Ref.~\onlinecite{Wachowicz:01}}
\footnotetext[13]{GGA, Ref.~\onlinecite{Wachowicz:01}} \footnotetext[14]{LDA,
Ref.~\onlinecite{Skriver:prb92}}
\end{table}

Table~\ref{clean_metal_wfs} lists the calculated work functions of the clean (111) surfaces (for Mg
the (0001) surface), compared to experimental values and values obtained in previous calculations.
Our results have been obtained using slabs consisting of six metal layers. A $25\times25$
$\mathbf{k}$-point sampling of the SBZ is applied, while allowing the top two metal layers to
relax. The GGA values generally are within $\sim 0.1$ eV of the experimental values. LDA still gives an acceptable accuracy, but tends to overestimate the work function somewhat. Only for Al(111) LDA gives a better value than GGA, as compared to experiment. Our results also agree with those obtained in previous computational studies; the small differences can be attributed to differences in the computational parameters, such as the functional, the basis set, and the lattice parameter.

\subsection{Structure of adsorbed monolayers}\label{sec:structures}

\begin{figure}[!tbp]
\includegraphics[width=7cm,clip=true]{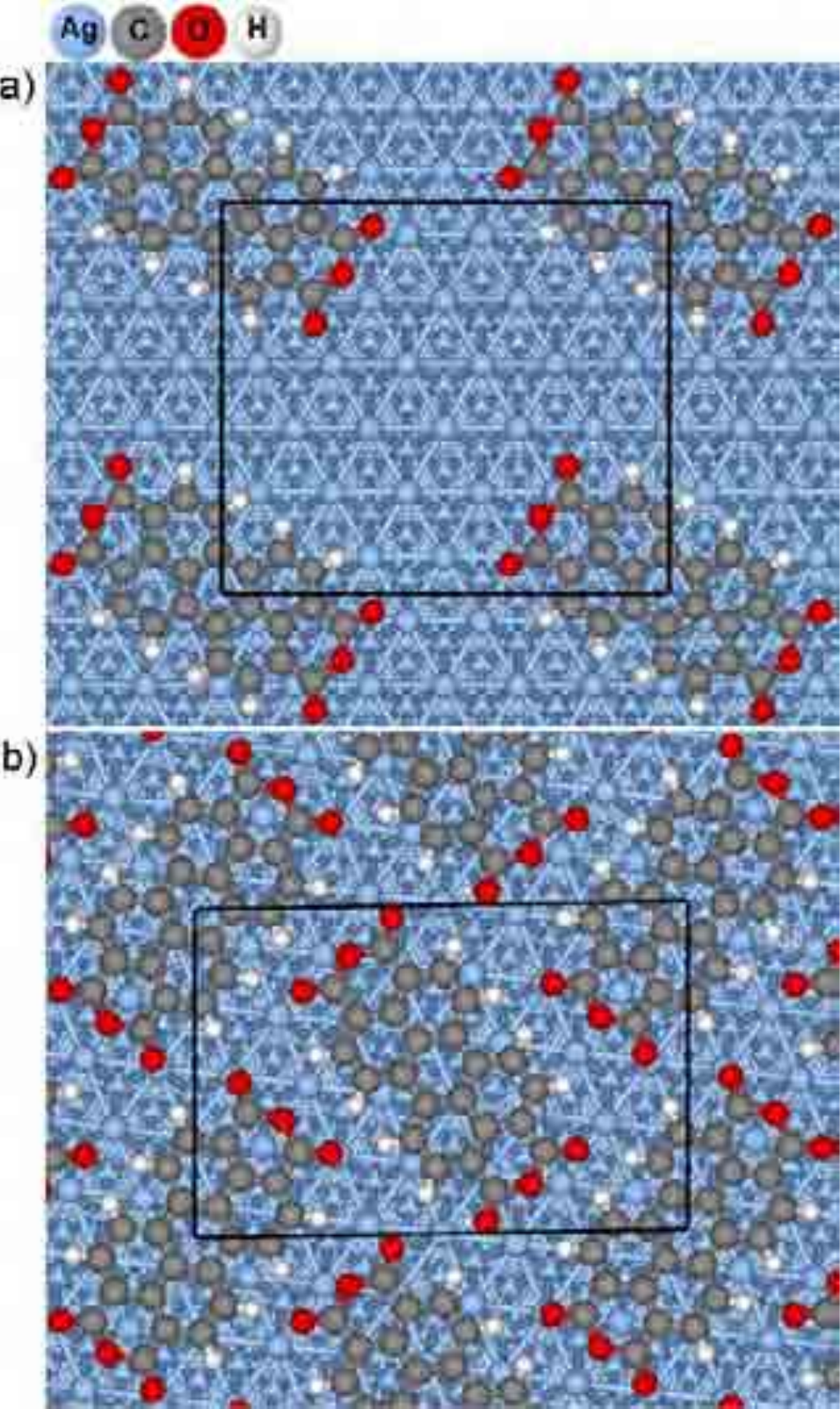}
\caption{(Color online) PTCDA monolayers on the Ag(111) surface. The rectangles denote the surface unit cells used
in calculations; (a) the dilute structure (with area $A=268$\AA$^2$); (b) the herringbone
structure ($A=243$\AA$^2$).} \label{structures}
\end{figure}

PTCDA monolayers adsorbed on Ag(111) and Au(111) surfaces have been studied in detail
experimentally.\cite{Umbach:surf_sci,Eremtchenko:nat03,Eremtchenko:njp04,Hauschild:prl05,Zou:ss06,Kraft:prb06,Temirov:nat06,Gerlach:prb07,Schmitz:prb97,Fenter:prb97,Nicoara:orgel06}
In a close-packed monolayer the PTCDA molecules lie flat on the surface in a ``herringbone''
structure with the centers of the PTCDA molecules located on surface bridge sites.\cite{Eremtchenko:nat03} The surface unit cell contains two PTCDA molecules, see
Fig.~\ref{structures}(b). The experimental distances between the carbon rings of the molecules and
the surface atoms are 2.86 \AA\ and 3.27 \AA\ for adsorption on Ag(111) and Au(111), respectively.\cite{Krause:jcp03,Gerlach:prb07} Experiments indicate a weak interaction between PTCDA and Au(111), consistent with physisorption,\cite{Eremtchenko:njp04,Nicoara:orgel06} and a somewhat stronger interaction between PTCDA and Ag(111).\cite{Eremtchenko:nat03,Eremtchenko:njp04,Hauschild:prl05,Zou:ss06,Kraft:prb06} PTCDA binds more strongly to open Ag surfaces and to surface steps.\cite{Du:prl06} We do not know of any such detailed studies on the structure of PTCDA adsorbed on the other metal (111) surfaces. We refrain from comparing our results to experiments where metals are deposited onto thin films of PTCDA, as this often leads to interdiffusion, which complicates the interpretation.\cite{Hirose:prb96,Kera:prb01,Kampen:ass04,Gavrila:apl06,Fuentes:apa06}

In our calculations we use the herringbone structure of PTCDA on Ag shown in
Fig.~\ref{structures}(b). The underlying Ag substrate contains 33 metal atoms per unit cell and our
supercell contains 175 atoms in total. Since the lattice parameters of Au, Ag and Al are similar,
see Table~\ref{lattice constants}, we use a similar supercell for PTCDA on these surfaces. For Mg
and Ca, we choose a herringbone structure that results in a packing density of PTCDA molecules
similar to that on the other surfaces. This results in 30 and 20 atoms per metal layer for Mg and
Ca, respectively.

To study the effect of the packing density of PTCDA molecules, we also perform calculations on the
structure used by Picozzi \emph{et al.},\cite{Picozzi:prb03} see Fig.~\ref{structures}(a). This
surface unit cell contains one PTCDA molecule. We refer to this structures as the ``dilute''
structure. The underlying Ag(111) surface then contains 36 atoms per unit cell, so that the
coverage of PTCDA molecules is $\sim \frac{1}{2}$ ML. The surface unit cells for the other metal substrates are chosen such, that the coverage remains close to this value. The distance between the PTCDA molecules is then sufficiently large for the molecules to have no direct interaction. As it is easier to vary the geometry of the molecule and substrate in the dilute structure, as compared to the close-packed herringbone structure, we use the former to study the energetics of PTCDA adsorption.
The optimized geometries of the PTCDA molecules in the two structures are very similar, as
demonstrated by Table~\ref{structural_data}, suggesting that close packing the molecules in the herringbone structure does not lead to a large intermolecular interaction. In addition, the GGA or LDA optimized geometries are very similar.

\begin{table}[btp]
\caption{Average bond lengths (in \AA) and bond angle of PTCDA adsorbed on Ag(111).} \label{structural_data}
\begin{ruledtabular}
\begin{tabular}{cccc}
   & LDA & GGA & GGA \\
   & dilute & dilute & herringbone  \\
\hline
 C$-$H               &1.10    &1.09   &1.09   \\
 C$-$C               &1.42    &1.43   &1.43  \\
 C$-$O\footnotemark[1]               &1.22    &1.23   &1.23   \\
 C$-$O\footnotemark[2]               &1.38    &1.40   &1.40   \\
 C$-$O$-$C$(^\circ)$     &125.2   &125.3  &124.7   \\
\end{tabular}
\end{ruledtabular}
\footnotetext[1]{carboxyl} \footnotetext[2]{anhydride}
\end{table}

Common DFT functionals describe strong chemical interactions well, but they fail to capture weaker (van der Waals) bonding correctly. If GGA functionals are used to
describe the adsorption of closed-shell molecules on metal surfaces, this can lead to underestimating the binding energy and overestimating the bond distance between molecule and surface. Using LDA functionals however can lead to a serious overbinding and a equilibrium distance that is too small.\cite{Morikawa:prb04,Morikawa:prb07} Previous GGA calculations of the binding energy of PTCDA on Ag(111) give slightly varying results, i.e. a moderate binding of $\sim 0.5$ eV/molecule, \cite{Du:prl06} or a very weak binding of $\lesssim 0.1$ eV/molecule, or even a purely repulsive binding curve.\cite{Picozzi:prb03,Ordejon:prl05,Hauschild:prl05a,Kraft:prb06,Romaner:njp09} In the calculations where binding was obtained, the equilibrium distance ($\sim 3.4$ \AA) is significantly larger than the experimental equilibrium distance ($2.9$ \AA).\cite{Hauschild:prl05,Gerlach:prb07} In contrast, LDA calculations on PTCDA on Ag(111) give equilibrium distances of 2.8 \AA\ \cite{Picozzi:prb03} and 2.7 \AA,\cite{Kraft:prb06} which are in better agreement with experiment. The LDA binding energy, $\sim 3$ eV/molecule, however, is suspiciously large.\cite{Kraft:prb06} We obtain very similar results in our GGA and LDA calculations for PTCDA on Ag(111).

\begin{figure}[btp]
\includegraphics[width=9cm,clip=true]{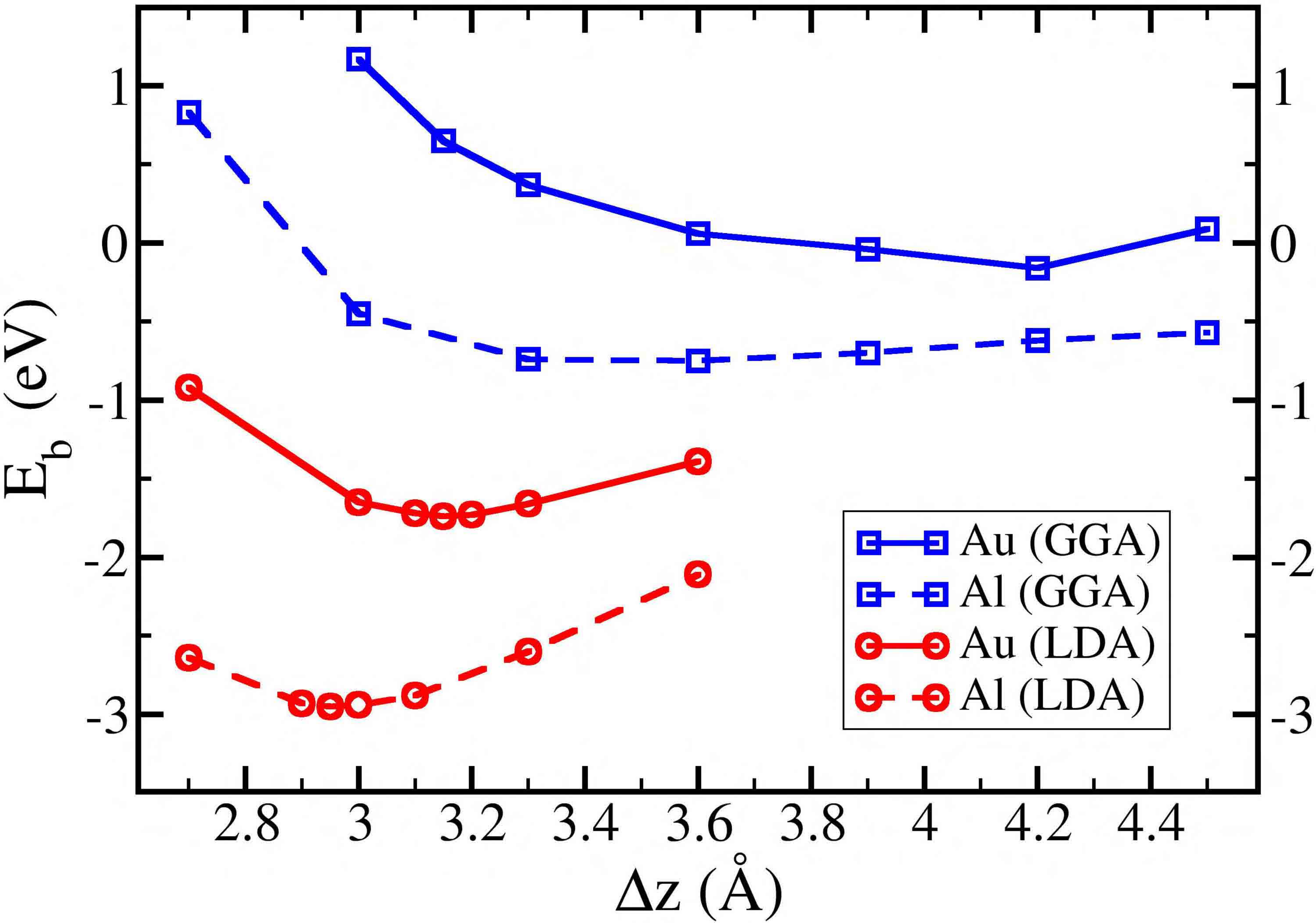}
\caption{(Color online) Binding energy curves for (planar) PTCDA on Au(111) (squares) and Al(111) (circles). GGA and LDA
values are indicated by the dashed and solid lines, respectively.} \label{binding_curves}
\end{figure}

To illustrate how general this trend is, Fig.~\ref{binding_curves} shows binding energy curves calculated with the dilute structure , where we varied the distance between the PTCDA molecule and the Au and Al(111) surfaces. The GGA results for PTCDA on Au lead to an extremely shallow binding curve with a minimum at a distance $> 4$ \AA\ and a very small binding energy of $\sim 0.1$ eV/molecule. Using GGA for PTCDA on Al gives a sizable binding energy ($\sim 0.8$ eV/molecule) and an equilibrium distance $\sim 3.5$ \AA. LDA calculations lead to much larger binding energies, i.e. 1.7 eV/molecule for PTCDA on Au and 3.0 eV/molecule for PTCDA on Al. The corresponding equilibrium distances are 3.15 \AA\ and 2.95 \AA, respectively. An LDA calculation for PTCDA on Ag gives an equilibrium distance of 2.75 \AA, which is somewhat smaller that the experimental value.

In conclusion, GGA and LDA give different results for the binding in weakly bonded
systems. GGA gives small molecule-surface binding energies and large equilibrium distances,
whereas LDA gives large binding energies and smaller equilibrium distances. As the reactivity of the surface increases along the series Au, Ag, Al, Mg and Ca, one expects the adsorption of PTCDA gradually changing from physisorption to chemisorption. The differences between the GGA and LDA become smaller and the results more reliable. For PTCDA on Mg(0001) and Ca(111) we obtain GGA binding energies of 2.3 and 8.4 eV/molecule, respectively, indicative of chemical bonding.

\begin{figure}[btp]
\includegraphics[width=7cm,clip=true]{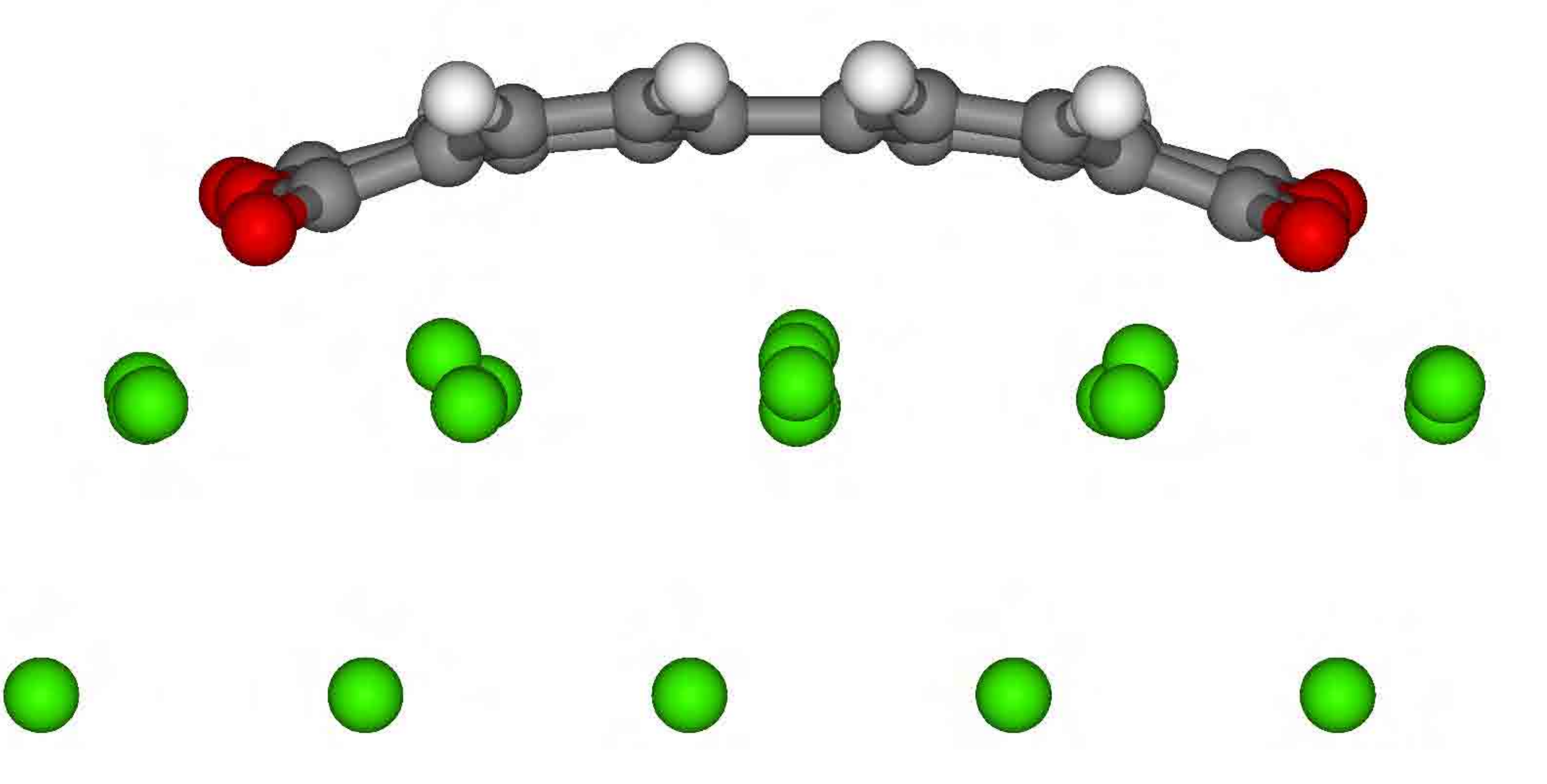}
\caption{(Color online) Optimized geometry of the arching PTCDA molecule on the Ca(111) surface.} \label{Ca_bent}
\end{figure}

We also find that, if the binding energy increases, the PTCDA molecule looses its
planar geometry. If the interaction between the PTCDA molecule and the surface is large, the molecule arches as shown in Fig.~\ref{Ca_bent}. The extend of this geometry deformation depends upon the metal substrate. Optimizing the geometry of PTCDA on Ca with GGA and defining the position of the surface by the average $z$ coordinate of the top layer of Ca atoms, we find that the outer carbon atoms of the perylene core are 0.8 \AA\ closer to the surface than the carbon atoms in the center. The latter have a distance of 2.6 \AA\ to the surface. Such short distances are indicative of a strong interaction between molecule and surface. A geometry deformation also occurs in the end groups of the PTCDA molecule, where the carboxyl oxygens move towards the surface, and the
anhydride oxygen moves way from the surface. A strong deformation of the
PTCDA molecule is accompanied by a rumpling of the surface, where metal atoms are lifted out of the surface, decreasing the distance with the molecule. For instance, the distances between the
carboxyl oxygens and the nearest Ca atoms are 2.3 \AA. Such short distances suggest the formation
of bonds, which matches the large binding energy of 8.4 eV/molecule.

The deformation of the adsorbed PTCDA molecule and that of the metal substrate decrease through the series Ca, Mg, Al, Ag, and Au, and is accompanied by a decrease of the binding energy. The deformation pattern of the PTCDA molecule qualitatively remains the same along this series, but the amplitude of the deformation decreases. For instance, the (GGA) distances between the carboxyl oxygens and the nearest Mg atoms are 2.4 \AA\, and the outer carbon and core carbon atoms are at 2.6 and 2.8 \AA\ from the surface, respectively. This matches the binding energy of 2.3 eV/molecule, which is much less than between PTCDA and Ca(111). Throughout the series Al, Ag and Au the binding energy, as well as the geometric deformation of the molecule, decrease monotonically.

For instance, if we fix the overall PTCDA-Ag(111) distance at 2.75 \AA\ and optimize the geometry, the molecule is only slightly arched. The outer carbon atoms are 0.1 \AA\ closer to the surface than the core carbon atoms. The carboxyl oxygen atoms are 0.2 \AA\ closer to the surface, whereas the anhydride oxygens are at approximately the same height as the core carbon atoms. This pattern is in fair agreement with experimental observations.\cite{Hauschild:prl05} At the end of the series, i.e. for PTCDA on Au(111), the binding energy is vanishingly small and the molecule and surface
have an undistorted, planar geometry. The binding energies and geometries of PTCDA on Al and Ag obtained in previous calculations follow the trends discussed above.\cite{Picozzi:prb03,Ordejon:prl05,Hauschild:prl05a,Romaner:njp09}

\subsection{Work functions}\label{sec:PTCDA_wfs}

From the calculations discussed in the previous section we can understand the trends in the
bonding and in the geometry of PTCDA adsorbed on the different metal surfaces. In this section we study the work function changes induced by the adsorption, which originate from a redistribution of charge at the MOI. The charge distribution is mainly determined by electrostatic and short-range exchange correlation interactions, which are represented well by LDA and GGA functionals.\cite{Rydberg:prl03,Thonhauser:prb07} For example, the charge transfer between the weakly interacting molecules in a charge transfer crystal is described well by LDA,\cite{Brocks:prb97} as is the work function graphene adsorbed on a metal surface.\cite{Giovannetti:prl08,Khomyakov:prb09} Given the bonding geometry of the molecule and surface, we expect therefore these functionals to give good values for the work function of molecular monolayers adsorbed on a metal surface. However, since they do not incorporate van der Waals interactions correctly, calculations lead to a considerable uncertainty in the molecule-surface equilibrium distance for weakly bonded systems, as discussed in the previous section. Because of this uncertainty we investigate the formation of interface dipoles in a number of steps. We start with the dilute structure and perform calculations for fixed molecule surface distances $d$ in the range 3.0-3.6 \AA. Results obtained with GGA and with LDA are then compared. In the second step we switch to the more densely packed herringbone structure that is observed experimentally, which allows us to study the effect of the packing density. Finally, we discuss the effects of full geometry relaxation of the molecules and the surface.

\begin{figure}[!tbp]
\includegraphics[width=8cm,clip=true]{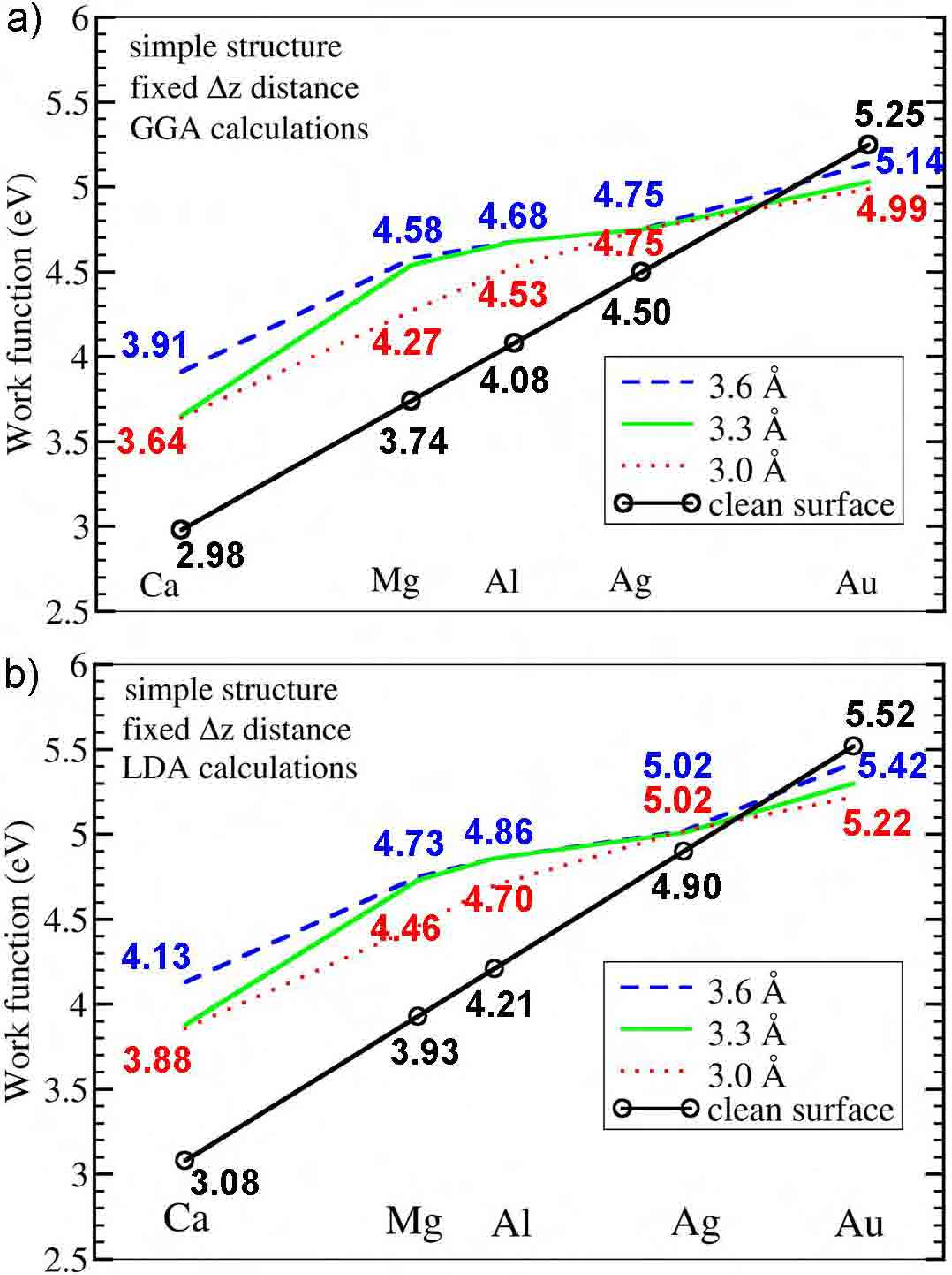}
\caption{(Color online) Work functions of a PTCDA monolayer on (111) metal surfaces [(0001) for Mg] in the dilute
structure, calculated using (a) the GGA and (b) the LDA functionals. The clean metal work function
is given along the $x$-axis. The numbers give the calculated values, the lines guide the eye. The
bottom (black) curve refers to the clean metal surfaces. The top three curves are for different
distances $d$ between the PTCDA molecules and the surfaces.} \label{results_LDAGGA}
\end{figure}

Fig.~\ref{results_LDAGGA}(a) shows the work functions for a layer of planar PTCDA molecules adsorbed in the dilute structure on the different metal surfaces, calculated using the GGA functional. One immediate observation is that adsorption on Au(111) leads to a lowering of the work function as compared to the clean surface, whereas adsorption on the other metal surfaces leads to an increase of the work function. There is some dependence of the work function on the distance between the molecule and the surface, but it is not excessively large. By fitting a straight line through the curves in Fig.~\ref{results_LDAGGA}(a) one obtains (see Eq.~(\ref{S_parameter})) $S=0.5$ at $d=3.6$ \AA, and $S=0.6$ at $d=3.0$ \AA. These values are considerably lower than the $S=1$ that follows from the Schottky-Mott rule, indicating that significant interface dipoles are formed upon adsorption. Since the work function changes decrease somewhat upon decreasing the molecule-surface distance, the interface dipoles decrease with decreasing distance. The calculated values for $S$ are much higher than the $S\approx 0$ obtained experimentally \cite{Kahn:jpsb03,Hill:apl98}. We will show below that this discrepancy is resolved by increasing the packing density of the PTCDA molecules, which is only $\sim \frac{1}{2}$ ML in the dilute structure.

To illustrate the effect of using a different functional, Fig.~\ref{results_LDAGGA}(b) gives the
work functions for the dilute structure at fixed distances between the molecule and the
substrates, calculated using the LDA functional. Compared to the GGA results of
Fig.~\ref{results_LDAGGA}(a), the LDA work functions are generally somewhat higher, as was also
the case for the clean metal surfaces, see Table~\ref{clean_metal_wfs}. The changes in the work
functions upon adsorption calculated with LDA or GGA, are comparable however at the same
molecule-surface distance. It means that, although the binding between the molecules and the
surfaces calculated with LDA or GGA can be considerably different, as discussed in the previous
section, the charge redistribution upon adsorption is similar, if we consider the same
molecule-surface distance. Since the GGA work functions of the clean metal surfaces are somewhat
closer to the experimental values, see Table~\ref{clean_metal_wfs}, we will use GGA values
for the adsorbed layers in the following.

\begin{figure}[!tbp]
\includegraphics[width=8cm,clip=true]{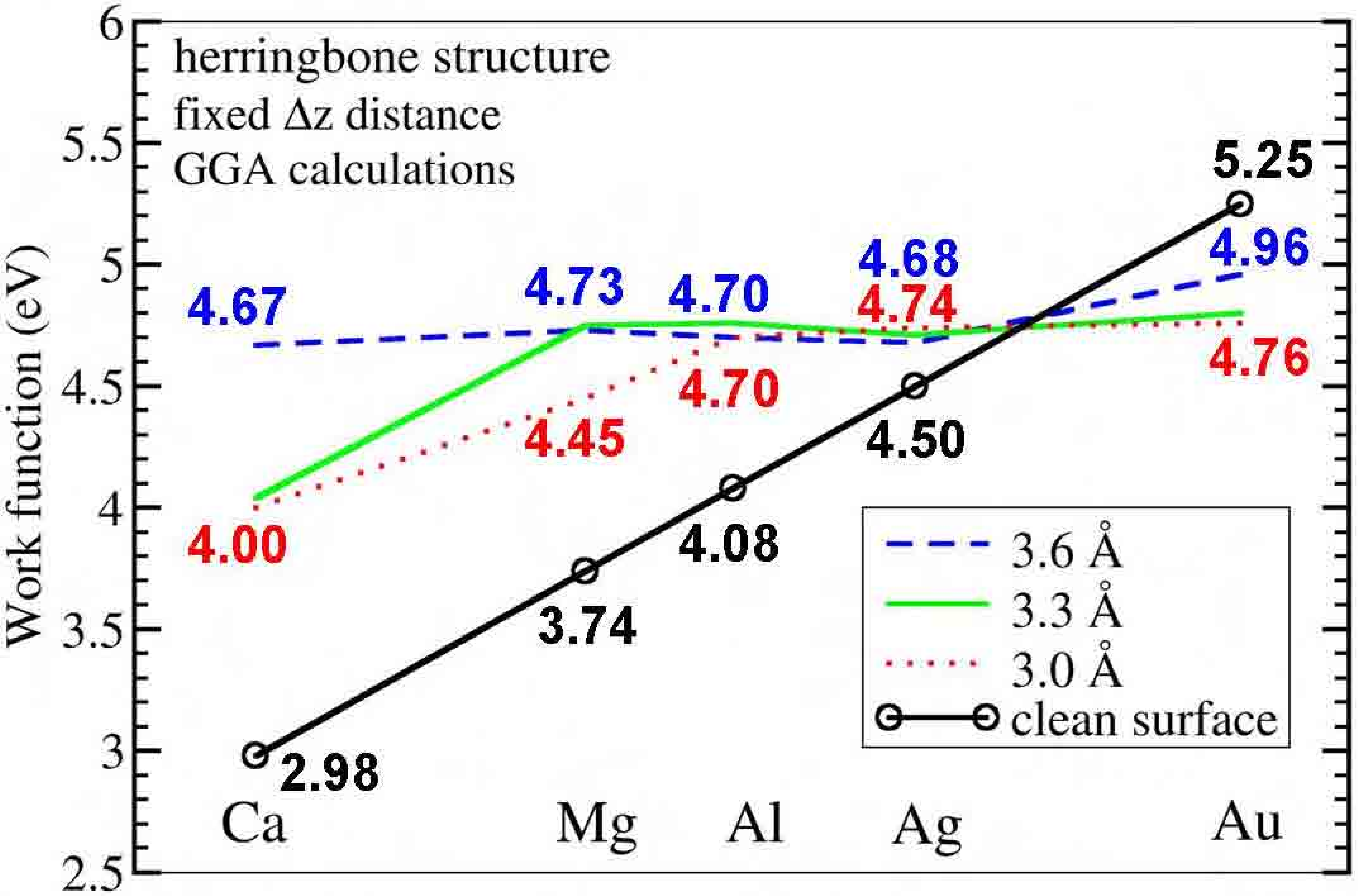}
\caption{(Color online) As Fig.~\ref{results_LDAGGA} for PTCDA in the herringbone structure}
\label{results_herringbone}
\end{figure}

Fig.~\ref{results_herringbone} shows the work functions of the herringbone structure of PTCDA on
metal surfaces, calculated using the GGA functional. As for the dilute structure,
Fig.~\ref{results_LDAGGA}(a), adsorption on Au(111) lowers the work function, and on the other
metal surfaces it increases the work function. In the herringbone structure the work function
shifts are much larger however. Most strikingly, as one can observe in
Fig.~\ref{results_herringbone}, the work function is pinned at $\sim
4.7$ eV over a considerable range of metal substrates and molecule-surface distances. The pinning
leads to $S=0$, which is in agreement with experiment. Moreover, the value of the pinned work
function is close to the value found experimentally \cite{Kahn:jpsb03,Hill:apl98}. Deviations from pinning are observed only for low work function metals and short molecule-surface distances, i.e.
$d \leq 3.3$ \AA\ for Ca and $d = 3.0$ \AA\ for Mg.

LDA gives very short molecule-surface distances, which can serve as lower bounds $d_{\min}$. We have optimized the geometries at $d_{\min}$ of an adsorbed PTCDA monolayer in the herringbone structure and calculated the work functions with GGA, see Table~\ref{tab:wfPTCDA}. These work function results for adsorption on Au, Ag and Al(111) are very close to the pinning value in Fig.~\ref{results_herringbone}. The work functions for adsorption on Mg(0001) and Ca(111) drop to a lower value. As discussed in the previous section, the Ca and Mg surfaces become somewhat unstable at $d_{\min}$, i.e. metal atoms are pulled up from the surface, which is accompanied by a strong arching distortion of the PTCDA molecule. Such a deformation generates a molecular dipole moment perpendicular to the surface that decreases the work function.\cite{Romaner:njp09}

\begin{table}[btp]
\caption{LDA optimized molecule-surface distances $d_{\min}$ and GGA work functions $W$ of PTCDA monolayers in the herringbone structure, adsorbed on metal(111) (for Mg (0001)) surfaces.} \label{tab:wfPTCDA}
\begin{ruledtabular}
\begin{tabular}{ccc}
   & $d_{\min}$ \AA & $W$ eV \\
\hline
 Au   & 3.15 & 4.78 \\
 Ag   & 2.75 & 4.65 \\
 Al   & 2.95 & 4.68 \\
 Mg   & 2.7-2.8 & 3.80 \\
 Ca   & 2.3-2.4 & 3.39 \\
\end{tabular}
\end{ruledtabular}
\end{table}

\section{Benzene and perylene}\label{benzene_and_perylene}

Although there is some spread in the numbers measured, experiments clearly indicate a lowering of the work function after adsorption of a benzene monolayer on several metal surfaces. Measured work function shifts are 0.18 eV on Al(111),\cite{Duschek:cpl00} 0.3 eV and 0.7 eV on Ag(111),\cite{Dudde:ss90,Zhou:ss90,Gaffney:cp00} 0.7 eV and 1.05 eV on Cu(111),\cite{Velic:jcp98,Bagus:apl05} and 1.10 eV on Au(111).\cite{Bagus:apl05} Previous GGA calculations for benzene on Al(111) gave an equilibrium distance of 3.7-3.8 \AA.\cite{Duschek:cpl00} Quantum chemical MP2 calculations for a single benzene molecule
adsorbed on a cluster of metal atoms gave equilibrium distances of 3.8 \AA\ for benzene on Au(111) and 4.0 \AA\ for benzene on Cu(111) \cite{Bagus:apl05}. We calculate the work functions of a
benzene monolayer adsorbed on different metal surfaces at a set of fixed distances in the same way as discussed in the previous section. A $(\sqrt{7} \times \sqrt{7})R19.1^\mathrm{o}$
structure is used as in Ref.~\onlinecite{Duschek:cpl00}, which gives a somewhat less than close-packed coverage of the surfaces.

\begin{figure}[!tbp]
\includegraphics[width=8cm,clip=true]{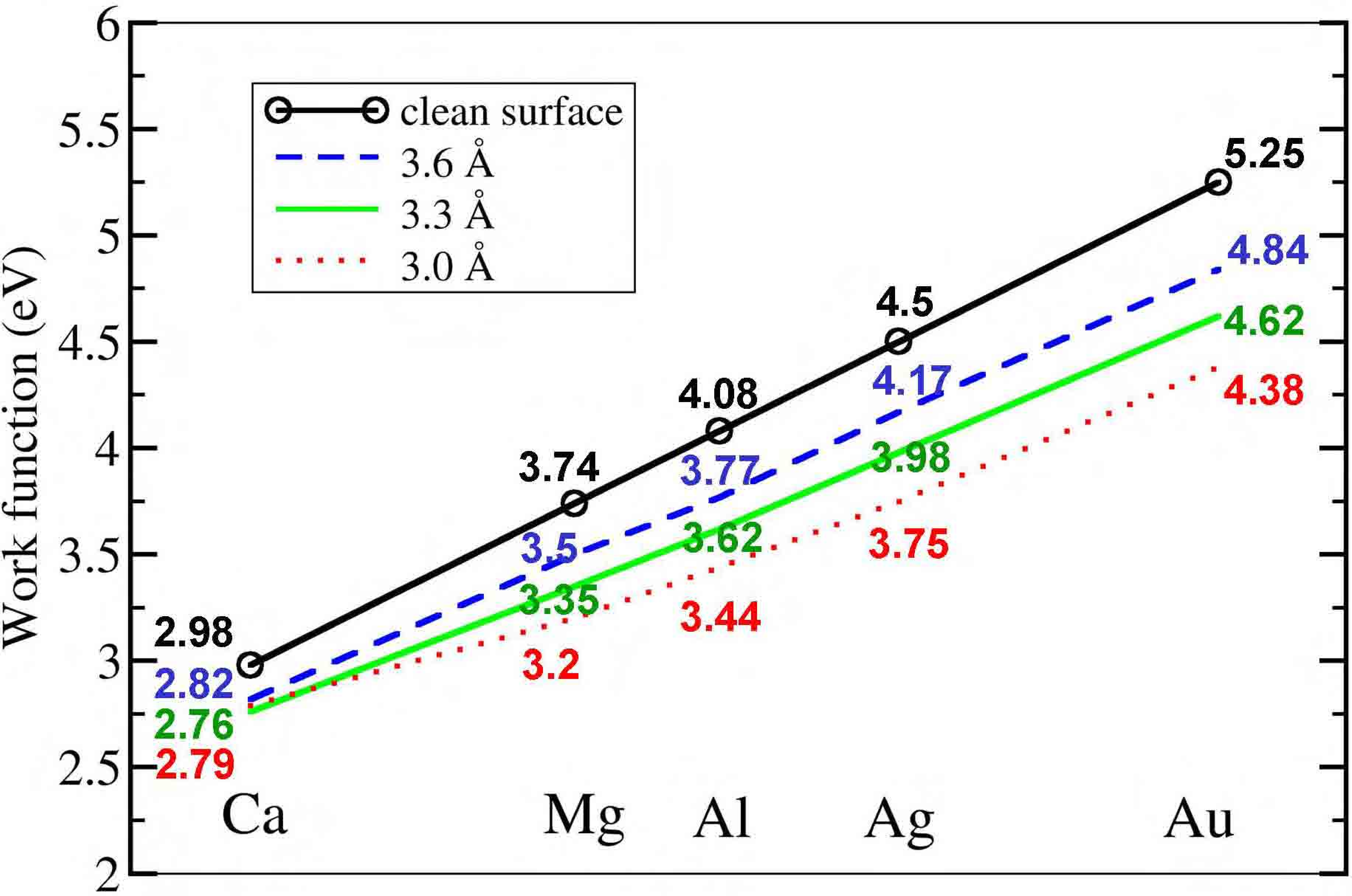}
\caption{(Color online) As Fig.~\ref{results_LDAGGA} for benzene in a $(\sqrt{7} \times
\sqrt{7})R19.1^\mathrm{o}$ structure.\cite{Duschek:cpl00}} \label{results_benzene}
\end{figure}

The (GGA) results are given in Fig.~\ref{results_benzene}. The most important observation is that
adsorption of benzene leads to a decrease of the work function for all the surfaces studied. This
is very different from the effect of PTCDA adsorption, see Figs.~\ref{results_LDAGGA} and
\ref{results_herringbone}. Moreover, adsorption of benzene gives a work function lowering that is
of a similar size for all surfaces (at a fixed molecule-surface distance). This leads to $S=0.9$ at $d=3.6$ \AA, and $S=0.8$ at $d=3.0$ \AA, which is not extremely far from the Schottky-Mott limit $S=1$. The absolute size of the work function shift depends on the molecule-surface distance with the number at $d=3.0$ \AA\ being roughly twice as large as that at $d=3.6$ \AA. The sign of the work function shift, its relatively weak dependence on the metal, and its sensitivity to the molecule-metal distance all point to an interpretation in terms of the pillow effect. The effect is determined by the Pauli repulsion between the molecular and surface electrons, which decreases the surface dipole and therefore the work function.\cite{Bagus:prl02,Dasilva:prl03,Bagus:apl05} Pauli repulsion critically depends on the overlap between the molecular and surface wave functions and therefore on the distance between the molecule and the surface. Our calculations and previous
calculations \cite{Duschek:cpl00,Bagus:apl05} suggest that the distances between the benzene
molecule and the metal surfaces are rather large, i.e. $\gtrsim 3.5$ \AA.

Summarizing, adsorption of the large PTCDA molecule leads to work function pinning ($S \approx
0$), and adsorption of the small benzene molecule gives a uniform work function lowering ($S \approx 1$). It is then interesting to study the adsorption of an intermediately sized molecule, such as perylene. The structure of a perylene monolayer on a metal surface is less well-established than that of a PTCDA monolayer. A herringbone structure similar to PTCDA is the structures proposed for a close-packed perylene monolayer on Ag(111) and Au(111).\cite{Eremtchenko:njp04,Seidel:prb01,Eremtchenko:jmatres04} UPS measurements give a decrease of
the work function of perylene on Au and Ag by 0.8 eV and 0.6 eV, respectively, and an increase of
the work function of perylene on Ca by 0.3 eV.\cite{Yan:apl02}

\begin{figure}[!tbp]
\includegraphics[width=8cm,clip=true]{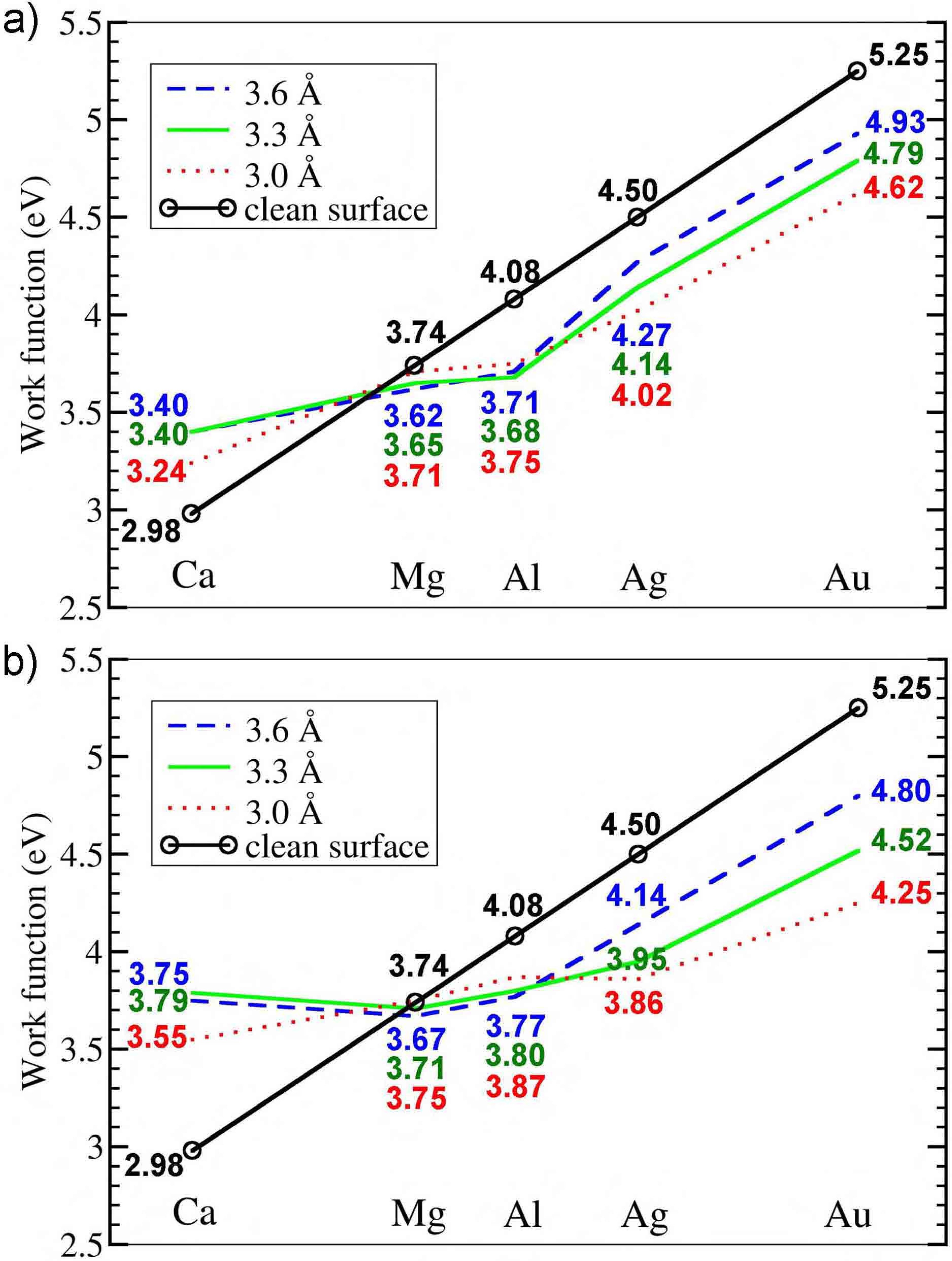}
\caption{(Color online) As Fig.~\ref{results_LDAGGA} for perylene in (a) the dilute structure and (b) the
herringbone structure.} \label{results_perylene}
\end{figure}

We perform calculations for a perylene monolayer adsorbed on different metal surfaces at a set of
fixed distances. Analogous to PTCDA we use two different structures, i.e. a close-packed
herringbone structure, and a dilute structure with a packing density of $\sim \frac{1}{2}$ ML. The calculated work functions are shown in Fig.~\ref{results_perylene}. Two regimes can be
distinguished. For the high work function surfaces (Au, Ag), adsorption of perylene leads to a
lowering of the work function, whereas for the low work function surface of Ca, adsorption of
perylene increases the work function. These results are in qualitative agreement with experiment. The curves in Fig.~\ref{results_perylene} indicate that the transition between these two regimes takes place in the range Mg-Al.

From these curves the transition between the two regimes can be quantified. Starting with the
results obtained for the dilute structure at a molecule-surface distance $d=3.6$ \AA, a line
through the points for Ca, Mg and Al gives $S=0.3$, whereas a line through the points for Al, Ag
and Au leads to $S=1.0$. For the herringbone structure the same procedure for $d=3.6$ \AA\ gives
$S=0$ and $S=0.9$, respectively. It is instructive to compare these $S$ values to the values
obtained for benzene and PTCDA. It suggests that for Ca, Mg and Al one obtains pinning of the work function upon perylene absorption, similar to PTCDA, see Fig.~\ref{results_herringbone}, whereas
for Ag and Au one finds a uniform work function decrease, similar to benzene, see
Fig.~\ref{results_benzene}. Upon decreasing the distance between the perylene molecules and the
surfaces the $S$ values in the high and low work function regimes become somewhat closer and the
transition between the two regimes becomes less sharp. Note that the distance dependence in the low work function regime resembles the distance dependence of the PTCDA case, whereas in the high work function regime it resembles the benzene case. Experimentally such a transition is observed for Alq$_3$ adsorbed on different surfaces. For adsorption on low work function metals $S\approx 0$, whereas for adsorption on high work function metals $S\approx 1$.\cite{Tang:cpl04}

\section{Discussion}\label{discussion}

The results presented in Fig.~\ref{results_herringbone} show that adsorption of a PTCDA monolayer
pins the work functions at $\sim 4.7$ eV for a broad range of distances. The fact that one gets
pinning, as well as the value of the pinning level, are in good agreement with experimental
observations. The results obtained for benzene and perylene adsorption are in qualitative
agreement with available experimental results, i.e. adsorption of benzene leads to a lowering of
the work function in all cases, and adsorption of perylene gives a work function decrease for high work function metals and a work function increase for low work function metals. In this section we analyze this behavior and interpret the results. First we analyze the density of states, then we consider explicitly the charge transfer at the molecule-substrate interface, and finally we formulate a simple phenomenological model.

\subsection{Density of states}
\begin{figure}[btp]
\includegraphics[width=6.5cm,clip=true]{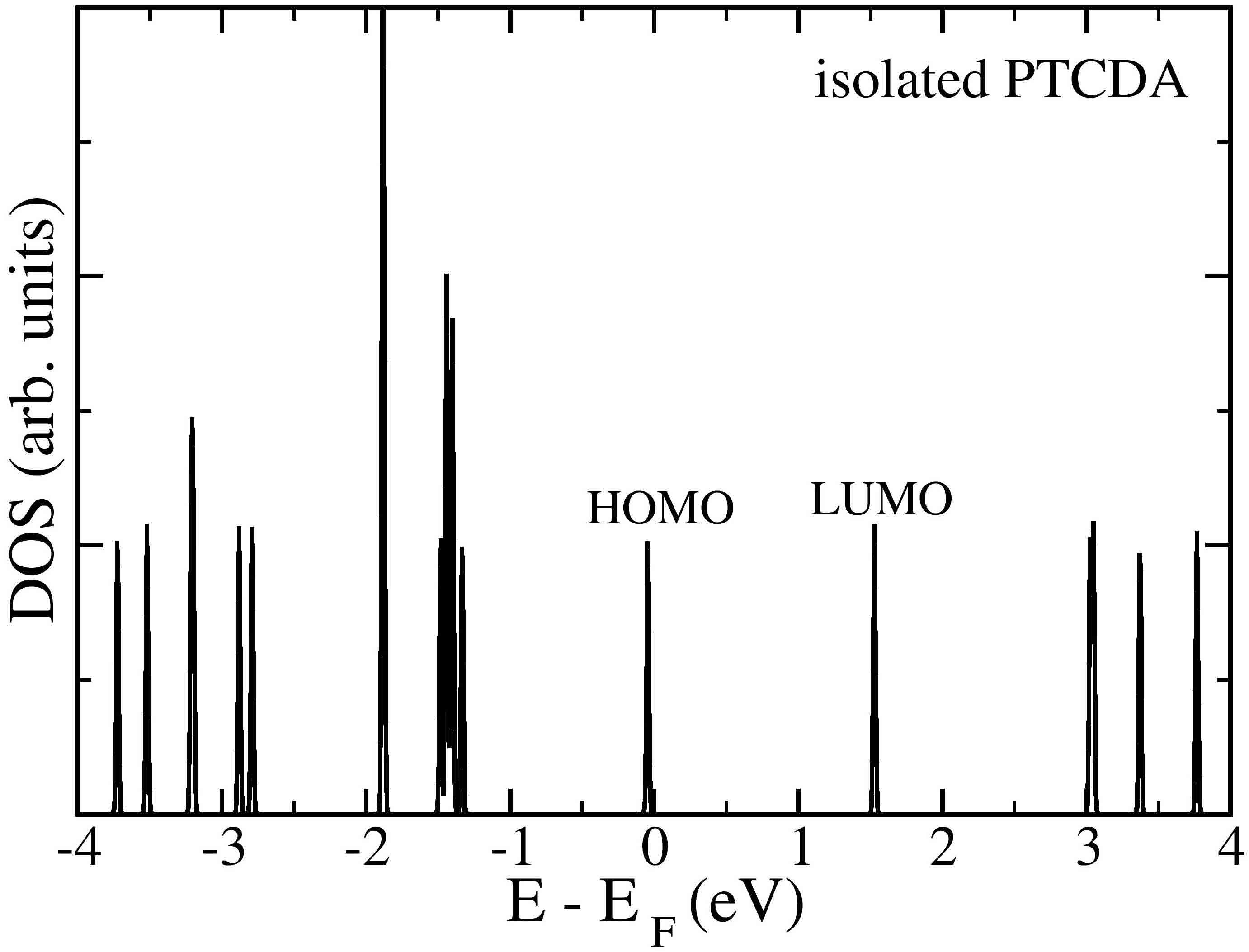}
\caption{Total DOS of an isolated PTCDA molecule calculated using a Gaussian broadening of 0.01 eV.
The highest occupied molecular orbital (HOMO) and the lowest unoccupied molecular orbital (LUMO)
are indicated. The energy is set to zero at the position of the HOMO. With respect to the vacuum
level the HOMO and LUMO levels are at $-6.24$ eV and $-4.64$ eV, respectively.} \label{LDOS_PTCDA}
\end{figure}

Figure~\ref{LDOS_PTCDA} gives the Kohn-Sham (KS) density of states (DOS) of an isolated PTCDA
molecule, calculated using the GGA functional. The energies of the highest occupied molecular
orbital (HOMO) and the lowest unoccupied molecular orbital (LUMO) are at $-6.24$ eV respectively
$-4.64$ eV with respect to the vacuum level. The KS spectrum is similar to that obtained in
previous DFT calculations.\cite{Vazquez:epl04,Kera:prb01,Vazquez:thesis,Dori:prb06} The GGA
HOMO-LUMO gap of 1.60 eV also agrees with the value found in other GGA calculations.\cite{Picozzi:prb03,Vazquez:epl04}

The interpretation of KS energy levels is the subject of a long-standing debate. From calculations with advanced functionals it is argued that the KS energies of all occupied molecular orbitals
correspond to vertical ionization potentials (IPs), as can be extracted from a photoemission spectrum, for instance. Approximative functionals such as GGA then still give a reasonable ionization spectrum, but it is shifted to a higher energy by approximately a constant.\cite{Chong:jcp02,Gritsenko:jcp03}  The work function is the lowest IP of an extended system. Even approximative functionals such as GGA or LDA usually give work functions that are
close to the experimental values, as is illustrated by Table~\ref{clean_metal_wfs}. Results of a
similar quality are obtained for work functions of adsorbed atomic and molecular layers.\cite{Rusu:prb06,Derenzi:prl05,Bredas:prl06,Rusu:jpcb06,Bredas:nanolett07,Rusu:jpcc07}

The KS energy levels corresponding to unoccupied molecular orbitals generally do not have such a simple interpretation. In particular, the energy of the DFT LUMO ($\epsilon_0$) should not correspond to the electron affinity (EA). From calculations with accurate functionals it is shown that $\epsilon_0<-\mathrm{EA}$, both for molecules,\cite{Gruenig:jcp02} as well as for extended systems.\cite{Gonze:prb04,Gruenig:jcp06} The difference between $|\epsilon_0|$ and the EA can be several eV's, which is attributed to the fact that an accurate functional has a discontinuous derivative as function of the number of electrons.\cite{Gonze:prb04,Gruenig:jcp06}

A similar difference is found for approximative continuous functionals such as GGA or LDA. However, it is well-known that for these approximative functionals, Slater's transition state approach allows for a simple estimate of the EA.\cite{Slater:74} We define $\epsilon_0$ as the KS energy of the LUMO of the neutral molecule, and $\epsilon_M(1)$ as the KS energy of the singly occupied HOMO of the ion that has one additional electron. The functionals allow for a fractional occupancy of the level $\epsilon_M$ with $N$ electrons, and using Janak's theorem\cite{Janak:prb78}
\begin{equation}
\epsilon_M = \frac{\partial E_\mathrm{tot}}{\partial N},
\end{equation}
where $E_\mathrm{tot}$ is the total energy, one can write
\begin{equation}
\mathrm{EA} = -\int_0^1\epsilon_M(N) dN.
\end{equation}
To a good approximation the dependence of $\epsilon_M$ on the occupancy is linear and can be parameterized as
\begin{equation}
\epsilon_M(N) = \epsilon_0 + U N, \label{eq:eps}
\end{equation}
where $U$ is the effective charging energy per electron.\cite{Sabin:ijqc00,fn:charging} This then gives
\begin{equation}
\mathrm{EA} = -\epsilon_0 - \frac{1}{2} U. \label{eq:ea}
\end{equation}
This procedure gives results that agree very well with charging energies for isolated conjugated molecules extracted from total energy calculations.\cite{Brocks:prl04}

From separate SCF calculations on the neutral molecule and the singly charged ion one can extract $\epsilon_0$ and $\epsilon_M(1)$, and calculate $U = \epsilon_M(1) - \epsilon_0$. Following this procedure, we extract a charging energy $U = 3.31$ eV from calculations on PTCDA$^0$ and PTCDA$^-$. Using the LUMO energy $\epsilon_0=-4.64$ eV, we then find from Eq.~(\ref{eq:ea}) $\mathrm{EA} = 2.98$ eV for PTCDA. This value is in very good agreement with the value $\mathrm{EA} = 2.96$ eV we extract from a $\Delta$SCF total energy difference calculation. Slater's transition state approach can also be used to calculate the IP. Assuming that the charging energy for holes on PTCDA is the same as for electrons and using the HOMO energy of $-6.24$ eV, then yields $\mathrm{IP} = 7.90$ eV, which is in fair agreement with the experimental PTCDA gas phase IP of $8.15$ eV cited in Ref.~\onlinecite{Vazquez:thesis}.

If a molecule is embedded in a crystal, its charging energy $U$ is reduced drastically because of
screening by the surrounding crystal.\cite{Tsiper:cpl02,Brocks:prl04} One can write
\begin{equation}
U = U_\mathrm{bare} - 2P^-, \label{eq:ueff}
\end{equation}
where $U_\mathrm{bare}$ is the charging energy of a bare molecule, and $P^-$ is the polarization energy associated with a singly charged molecular ion in a crystal.\cite{fn:pol} Using the polarization energy $P^- = 0.91$ eV from Ref.~\onlinecite{Tsiper:cpl02}, we obtain an effective charging energy of a PTCDA molecule embedded in a PTCDA crystal $U_\mathrm{cryst}=3.31-1.82=1.49$ eV. Slater's transition state model then gives $\mathrm{EA_{cryst}} = 3.90$ eV and $\mathrm{IP_{cryst}} = 6.99$ eV for the EA and IP of a PTCDA molecule in the crystal. This leads to a transport gap
$E_\mathrm{t}=\mathrm{IP_{cryst}}-\mathrm{EA_{cryst}}=3.09$ eV. The values of the IP and
$E_\mathrm{t}$ are in good agreement with the values extracted from experiment, i.e $6.7 \pm 0.2$
eV and $3.2 \pm 0.4$ eV, respectively.\cite{Hill:cpl00,Duhm:orgel08}

In summary, using the KS DOS as shown in Fig.~\ref{LDOS_PTCDA}, to calculate measurable quantities, one has to incorporate the charging energy $U$ of the molecule, cf. Eqs.~(\ref{eq:eps}) and (\ref{eq:ea}). $U$ strongly depends upon the interaction of the molecule with its environment. Screening by a metal substrate reduces $U$ significantly, for instance,\cite{Tsiper:cpl02,Hill:cpl00} and for electrons in completely delocalized states $U=0$.

\begin{figure}[btp]
\includegraphics[width=8.5cm,clip=true]{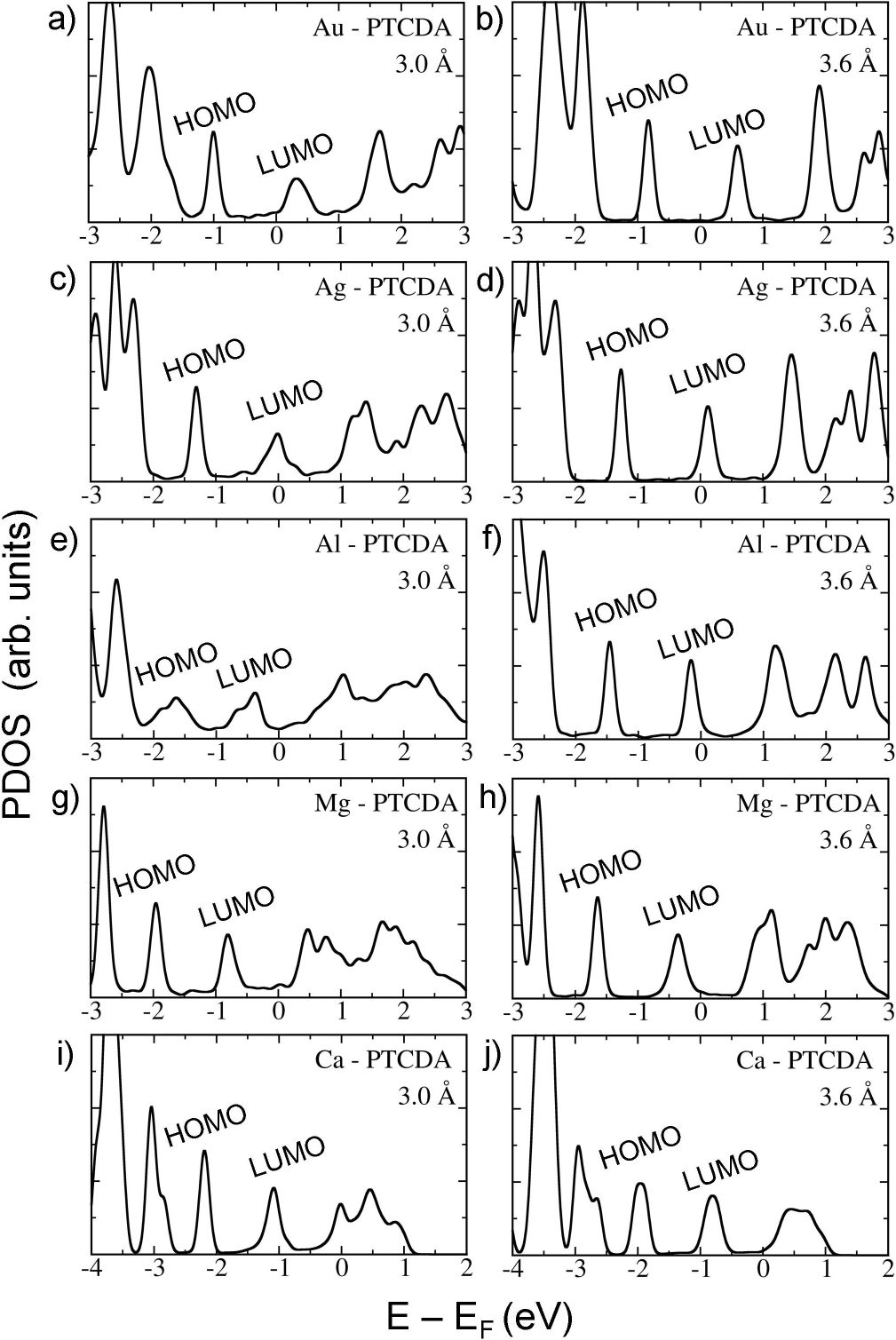}
\caption{Projected density of states (PDOS) of the PTCDA molecule adsorbed on metal surfaces at a
fixed distance $d$ in the herringbone structure, calculated using a Gaussian broadening of 0.1 eV;
left: $d=3.0$\AA; right: $d=3.6$\AA. The peaks corresponding to the molecular HOMO and LUMO levels
are labeled. } \label{DOS}
\end{figure}

We now consider the DOS of PTCDA monolayers adsorbed on metal surfaces. Since the full DOS is dominated by states originating from the metal substrate, we look at the DOS projected on the atoms of the molecule (PDOS), in order to identify the contributions of the molecular levels.
Fig.~\ref{DOS} gives the PDOSs PTCDA monolayers adsorbed on metal surfaces at distances of $d=3.0$ and $d=3.6$ \AA. We have applied a Gaussian broadening with a broadening parameter of 0.1 eV to avoid a spiky appearance of the PDOS. Comparison to Fig.~\ref{LDOS_PTCDA} allows us to
identify the the contributions of the molecular levels, in particular those states that are dominated by the molecular HOMO and LUMO. Although all states result from hybridization between molecular and metal states, it simplifies the analysis if we label them by their dominant molecular character.

The interaction between the molecule and the surfaces induces hybridization of molecular and metal states. For adsorption on simple, wide band metals this leads to a simple broadening of the molecular levels into resonances.\cite{Zangwill:book} For adsorbed PTCDA, the extend of this broadening is moderate. At $d=3.6$ \AA\ the typical width at half height of the HOMO and LUMO peaks is $\sim 0.2$ eV. The
widths increase with decreasing molecule-metal distances (to $\sim 0.3$ eV at $d=3.0$ \AA, for instance). These results agree with the widths found in previous DFT calculations of adsorbed PTCDA,\cite{Picozzi:prb03} as well as with those typically found for other adsorbed molecules such as pentacene.\cite{Morikawa:prb07} They are also in qualitative agreement with peak widths observed in scanning tunneling spectroscopy of PTCDA adsorbed on Ag and Au(111) surfaces.\cite{Kraft:prb06,Temirov:nat06,Nicoara:orgel06} The widths found in the calculations of Refs.~\onlinecite{Vazquez:epl04} and \onlinecite{Vazquez:ass04} are much larger.

Comparing the two columns in Fig.~\ref{DOS}, one observes that the PDOSs at the two distances are
qualitatively similar, but that the spectrum at $d=3.0$ \AA\ is shifted towards lower energy as
compared to the spectrum at $d=3.6$ \AA\ by  up to 0.5 eV, depending upon the metal substrate. This is caused by the pillow effect, as we will discuss in Sec.~\ref{sec:model}. The GGA HOMO-LUMO gap of PTCDA decreases somewhat upon adsorption, from 1.6 eV in the isolated molecule to $\sim 1.4$ eV (peak maximum to peak maximum) for PTCDA weakly interacting with Au(111) and $\sim 1.1$ eV for PTCDA strongly interacting with Ca(111).

One observation that can be made by comparing Figs.~\ref{results_herringbone} and
~\ref{DOS} is that work function pinning occurs when the Fermi level crosses the level of the
LUMO. For PTCDA on Au the LUMO is unoccupied, but already for PTCDA on Ag the LUMO gets
partially occupied. This implies that electron transfer takes place from the metal substrate to
the molecule. Judging from the upwards shift of the Fermi level along the columns of Fig.~\ref{DOS} the amount of electron transfer increases along the series Ag, Al, Mg, and Ca. As
long as the Fermi level is inside the LUMO peak, the work function is pinned, compare
Fig.~\ref{results_herringbone}. At the short molecule-surface distance of 3.0 \AA\ between PTCDA
and Ca, the Fermi level jumps to the next peak, i.e. the LUMO+1, see Fig.~\ref{DOS}(i). This is
accompanied by an ``unpinning'' of the work function, compare Fig.~\ref{results_herringbone}.

The most detailed experiments have been performed for PTCDA on Ag(111). In UPS and STM experiments a peak is observed at the Fermi level that is identified as the LUMO of the PTCDA molecule, whereas a peak at $\sim -1.6$ eV with respect to the Fermi level is labeled as the HOMO.\cite{Zou:ss06,Temirov:nat06,Kraft:prb06,Duhm:orgel08} In Fig.~\ref{DOS}(c), where the distance between PTCDA and the Ag surface is close to the experimental value, we find the HOMO at $-1.3$ eV and the LUMO at the Fermi level, which is in reasonable agreement with the experimental analysis. In STM and IPES experiments of PTCDA on Au(111) a peak is observed at 1.0-1.5 eV above the Fermi level that is associated with the EA level of the molecule.\cite{Tsiper:cpl02,Nicoara:orgel06,Kroeger:cpl07}. Applying Eqs.~(\ref{eq:ea}) and (\ref{eq:ueff}) puts the EA level at $P^-$ above the LUMO in Fig.~\ref{DOS}(a,b). Using the polarization energy $P^- = 0.97$ eV for a PTCDA ion adsorbed on a metal surface calculated in
Ref.~\onlinecite{Tsiper:cpl02} then brings the EA level in the experimentally observed range.

Pinning of the work function at MOIs has been interpreted in terms of a charge neutrality level
(CNL),\cite{Vazquez:epl04,Vazquez:ass04,Vazquez:jcp07} in analogy to Schottky barrier models for
conventional semiconductors.\cite{Flores:jpc77,Tersoff:prl84,Tersoff:prb85} The CNL model relies
on having a large continuum DOS at the metal-semiconductor interface, which fills the energy gap
of the semiconductor. The Fermi level is then pinned by these metal induced gap states (MIGS).\cite{Heine:phys_rev} Conventional semiconductors such as Si or GaAs have reactive surfaces with surface atoms carrying dangling bonds. The energies of these dangling bond states are within the semiconductor gap. Bonding at a metal-semiconductor interface leads to broadening of these states, which generates a large continuum DOS at the interface in the semiconductor gap.\cite{Louie:prb76,Lang:prb77} Closed shell molecules such as PTCDA do not have dangling bond
states within the HOMO-LUMO gap. The creation of a large DOS at a MOI within the HOMO-LUMO gap
then depends upon a large broadening of the molecular levels. We do not observe such a large
broadening.

\begin{figure}[btp]
\includegraphics[width=8.5cm,clip=true]{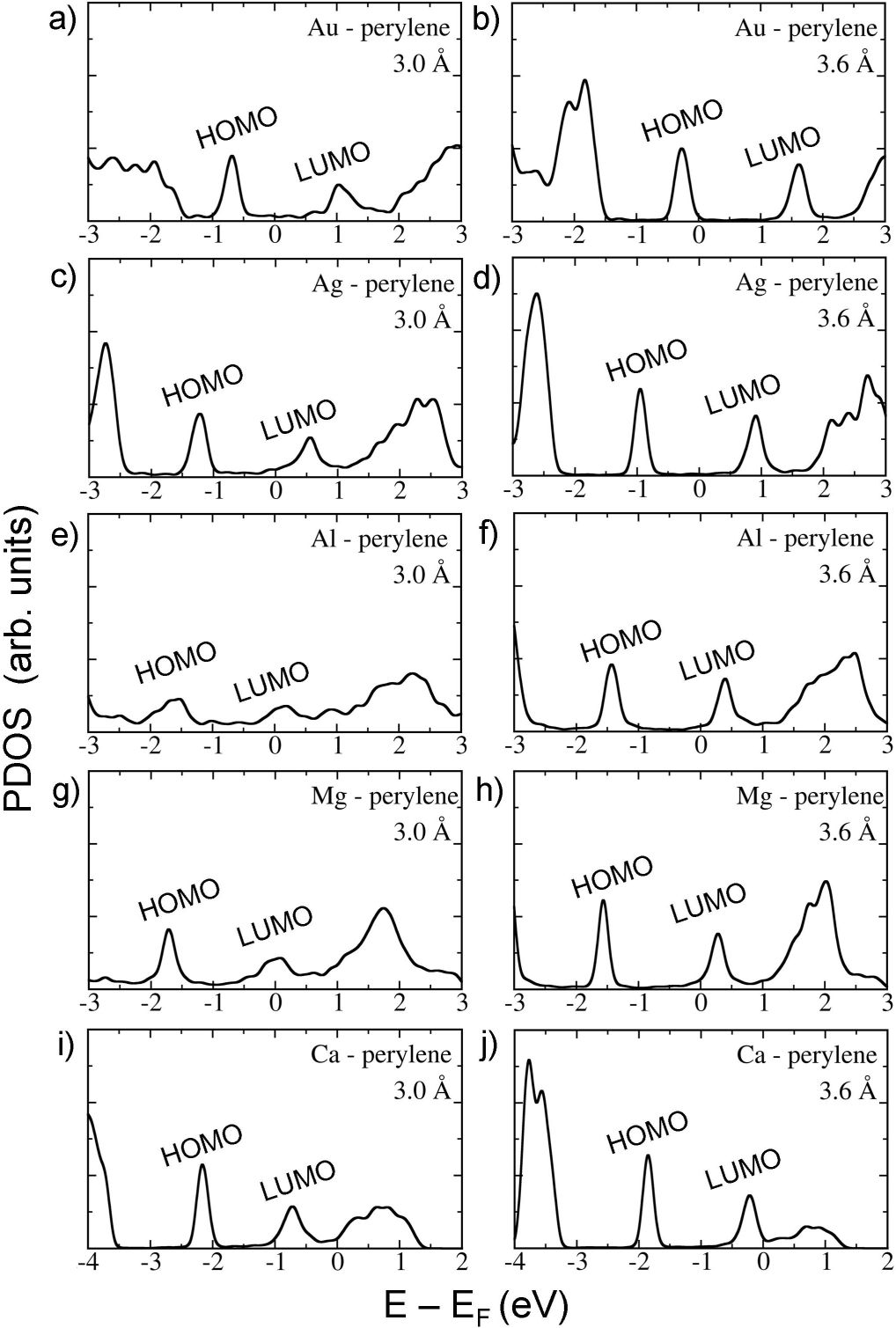}
\caption{As Fig.~\ref{DOS} for perylene adsorbed on metal surfaces. } \label{perylene_DOS}
\end{figure}

\begin{figure*}[btp]
\caption{(Color online) Laterally averaged electron density difference $\Delta \overline{n}(z)$ for PTCDA adsorbed
on (a) Ca(111), (b) Al(111), and (c) Au(111) at a fixed distance $d=3.0$ \AA. The $+/-$ indicate
the direction of the interface dipole. The charge on the molecule is estimated by integrating over
the shaded areas (see text). (d-f) Isodensity surface of $\Delta n(x,y,z)$ close to the molecular
plane.} \label{delta n}
\includegraphics[width=14cm,clip=true]{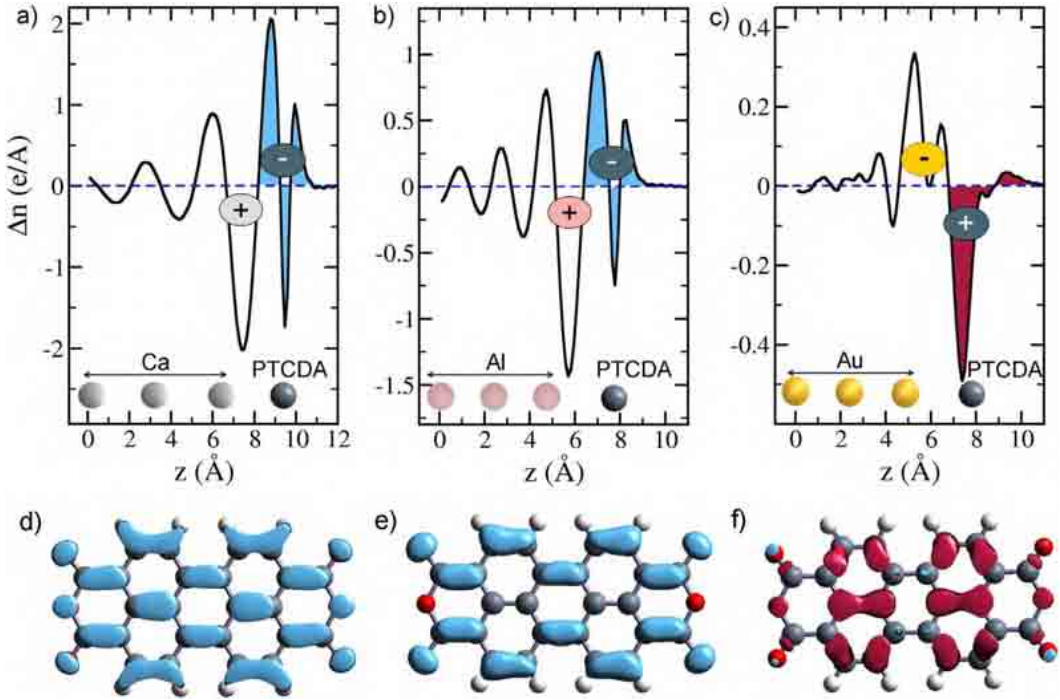}
\end{figure*}

The DOS of an isolated perylene molecule resembles that of PTCDA. The GGA HOMO-LUMO gap of 1.8 eV
is slightly larger than that of PTCDA. Fig.~\ref{perylene_DOS} gives the PDOS on a perylene
molecule adsorbed on metal surfaces at distances of $d=3.0$ and $d=3.6$ \AA. Comparison to
Fig.~\ref{results_perylene}(b) shows that pinning of the work function sets in as the Fermi level
reaches the LUMO. In other words, as for PTCDA, the work function is pinned by the LUMO of the
molecule. As the perylene LUMO lies $\sim 1$ eV higher than that of PTCDA, compare Figs.~\ref{DOS} and \ref{perylene_DOS}, the pinning level for adsorbed perylene is $\sim 3.7$ eV. For benzene we observe no pinning, see Fig.~\ref{results_benzene}, as the LUMO level is too high in energy and is not populated even if benzene is adsorbed on a low work function metal like Ca.

\subsection{Charge transfer and interface dipole}
The charge transfer at the PTCDA-metal interface can be visualized directly by calculating the
laterally averaged electron density difference
\begin{equation}
\Delta \overline{n}(z) =
\overline{n}_\mathrm{PTCDA/metal}(z)-\overline{n}_\mathrm{metal}(z)-\overline{n}_\mathrm{PTCDA}(z).
\label{eq:deltan}
\end{equation}
The electron density $\overline{n}_\mathrm{metal}$ and $\overline{n}_\mathrm{PTCDA}$ of the metal
substrate and the molecule are obtained in separate calculations with the substrate and the
molecule frozen in the adsorption geometry, using the same unit cell. The lateral averaging is
done as in Eq.~(\ref{eq:vav}). Examples of $\Delta \overline{n}(z)$ are shown in Fig.~\ref{delta
n}.

Fig.~\ref{delta n}(a-c) clearly show the formation of interface dipoles that are localized at the
PTCDA/metal interface. Note that the sign of the interface dipole moment of PTCDA on Au is
opposite to that of PTCDA on the other metal surfaces. For PTCDA on Au electrons are displaced
from the molecular region into the metal, whereas for PTCDA on other metals electrons are
displaced from the metal to the molecule. According to the PDOSs shown in Figs.~\ref{DOS}(c)-(j)
the latter can be interpreted as electron transfer from the metal to the LUMO of the molecule. In
contrast, the displacement of electrons for PTCDA on Au cannot straightforwardly be related to the
transfer of electrons to or from a molecular level, as can be observed from the PDOSs of
Figs.~\ref{DOS}(a),(b).

The charge displacement for PTCDA on Au can be interpreted in terms of the pillow effect.\cite{Bagus:prl02,Bagus:apl05} If the interaction between the molecule and the surface is not too
strong, then the wave function of the system can be written in good approximation as an
anti-symmetrized product of the wave functions of the separate molecule and the substrate.
Anti-symmetrization introduces the effects of exchange between the molecular and substrate
electrons, which leads to a repulsive interaction in case of a closed shell molecule, the
so-called Pauli repulsion. The electronic cloud of the molecule is usually ``harder'' than that of the metallic substrate, i.e. the former is less easily deformed. The net effect of Pauli repulsion is then that the electronic cloud that originally spilled out from the metal surface, is pushed
back into the metal by the molecular electronic cloud (as if the molecule lands on a pillow). The
push back of electrons into the metal can be observed in Fig.~\ref{delta n}(c). It lowers the
surface dipole, and therefore it lowers the work function.

One would expect that the spatial distribution of the displaced electrons reflect the shape of the
molecule. The pattern of electron depletion in the molecular region can be visualized by plotting
$\Delta n(x,y,z)$, as is shown in Fig.~\ref{delta n}(f). The nodal pattern roughly corresponds to
the second highest occupied molecular $\pi$ orbital of PTCDA (HOMO-5 in
Ref.~\onlinecite{Dori:prb06}). It confirms the conclusion drawn from the PDOSs of
Figs.~\ref{DOS}(a),(b) that the electron depletion is not due to a transfer of electrons from the
highest occupied state of the molecule to the Au surface.

The change in electron density for PTCDA on other metals is qualitatively different from that of
PTCDA on Au. Electrons are transferred from the metal to states of the molecule, which is clearly
demonstrated by plotting $\Delta n(x,y,z)$. The nodal pattern of $\Delta n(x,y,z)$ for PTCDA on
Al, Fig.~\ref{delta n}(e), corresponds to the LUMO of PTCDA, indicating that electrons are
transferred to this state, in agreement with Fig.~\ref{DOS}(d). For PTCDA on Ca, Fig.~\ref{delta
n}(e), the nodal pattern shows both features of the LUMO and of the LUMO+1.\cite{Kera:prb01} This
agrees with Fig.~\ref{DOS}(i), which shows that electrons are transferred to both these states.

One can calculate the interface dipole per adsorbed molecule as $\Delta \mu = eA \int z \Delta
\overline{n}(z) dz$, with $A$ the surface area of the adsorbed molecule. Alternatively the
interface dipole is extracted from the change in the work function $\Delta W$ upon adsorption of
the molecules \cite{Jackson:book}
\begin{equation}
\Delta \mu = \frac{\varepsilon_0 A}{e} \Delta W. \label{eq:dip}
\end{equation}
The results are given in Table~\ref{table:charge}. The total charge $q$ on the molecule can be
estimated by
\begin{equation}
q = -e \int_{z_0} \Delta \overline{n}(z) dz, \label{eq:charge}
\end{equation}
where $z_0$ is the point between the molecule and the surface where $\Delta \overline{n}(z_0)=0$.
The integration is indicated by shaded areas in Figs.~\ref{delta n}(a-c).

\begin{table}[btp]
\caption{Interface dipole per molecule $\Delta \mu$ (Eq.~(\ref{eq:dip})) and total molecular charge
$q$ (Eq.~(\ref{eq:charge}))for PTCDA on
metal surfaces. The distance between the molecules and the surface is fixed at 3.0 \AA.}
\label{table:charge}
\begin{ruledtabular}
\begin{tabular}{ccc}
 metal  & $\Delta \mu$ (D) & $q$ ($e$) \\
\hline
Au & $-1.58$ &  $+0.34$ \\
Ag &   0.77  &  $-0.31$ \\
Al &   2.00  &  $-0.82$ \\
Mg &   2.29  &  $-0.84$ \\
Ca &   3.29  &  $-1.34$ \\
\end{tabular}
\end{ruledtabular}
\end{table}

Comparing the values for Ag to Ca in Table~\ref{table:charge}, one notices an increase in the
interface dipole and in the number of electrons transferred from the metal surface to the
molecule. This is consistent with the change in the work function upon adsorption and with the
PDOSs, see Figs.~\ref{results_herringbone} and \ref{LDOS_PTCDA}. The charge distributions of the
interface dipole on Ag, Al, Mg, and Ca are similar. The distributions shown in Figs.~\ref{delta n}(a,b) can be interpreted as
electronic charge placed in a $\pi$-orbital on the PTCDA molecule, which is screened by electrons
in the metal, which leads to characteristic Friedel oscillations in the metal substrate.

Adsorption of PTCDA on Au is qualitatively different from adsorption on the other metal surfaces,
as can be judged from the signs of the interface dipoles and the charges. The charge distribution of PTCDA on Au, Fig.~\ref{delta n}(c), is also qualitatively
different. It is more localized in the region between the molecule and the surface, as can be
expected if it is due to Pauli repulsion (i.e. the pillow effect), since the latter is active in
the region where the molecular and the surface wave functions overlap.

Since this overlap decreases with increasing distance between the molecule and the surface, one
expects the charge displacement to decrease accordingly. Indeed the charge on PTCDA adsorbed on
Au, calculated using Eq.~(\ref{eq:charge}), at a distance $d=3.6$ \AA\ is 0.23$e$, as compared to
0.34$e$ at $d=3.0$ \AA, see Table~\ref{table:charge}. The pillow effect is a very general
mechanism that should be operative for any adsorbed closed shell molecule, even if electrons are
transferred from the substrate to the LUMO, as for PTCDA adsorbed on other metal surfaces. One
would expect that this leads to an interface dipole that depends strongly on the molecule-surface
distance. However, Fig.~\ref{results_herringbone} shows that for PTCDA on Ag and Al (and to a
lesser extend also for PTCDA on Mg), the work function, and therefore the interface dipole, is
independent of the distance over a considerable range.

\subsection{Model} \label{sec:model}
In this section we aim at setting up a simple model that explains qualitatively the most prominent features of the work functions of molecular monolayers adsorbed on metal surfaces. In particular, we want to identify the different regimes, i.e. work function decrease by the pillow effect and work function pinning by charge transfer. Moreover, the lack of the work function dependence on the molecule surface distance in the pinning regime should be clarified.

To construct this model we assume that the relevant energy scale is set by molecular properties such as the HOMO-LUMO gap, and neglect the broadening of molecular levels introduced by the interaction between the molecule and the surface. We assume that electrons can be transferred to the LUMO, whose energy $\epsilon_M(N)$ depends on the occupation number $N$, as in Eq.~(\ref{eq:eps}). In the spirit of Slater's transition state approach we assume that $N$ can take any value $0 \leq N \leq 2$.\cite{fn:ICT} The charging energy $U$ depends
upon the environment of the molecule. All molecules in a monolayer have the same occupation number and in the effective $U$ the interactions between all molecules should be taken into
account, as well as the interactions with the metal substrate.

\begin{figure}[btp]
\includegraphics[width=5.5cm,clip=true]{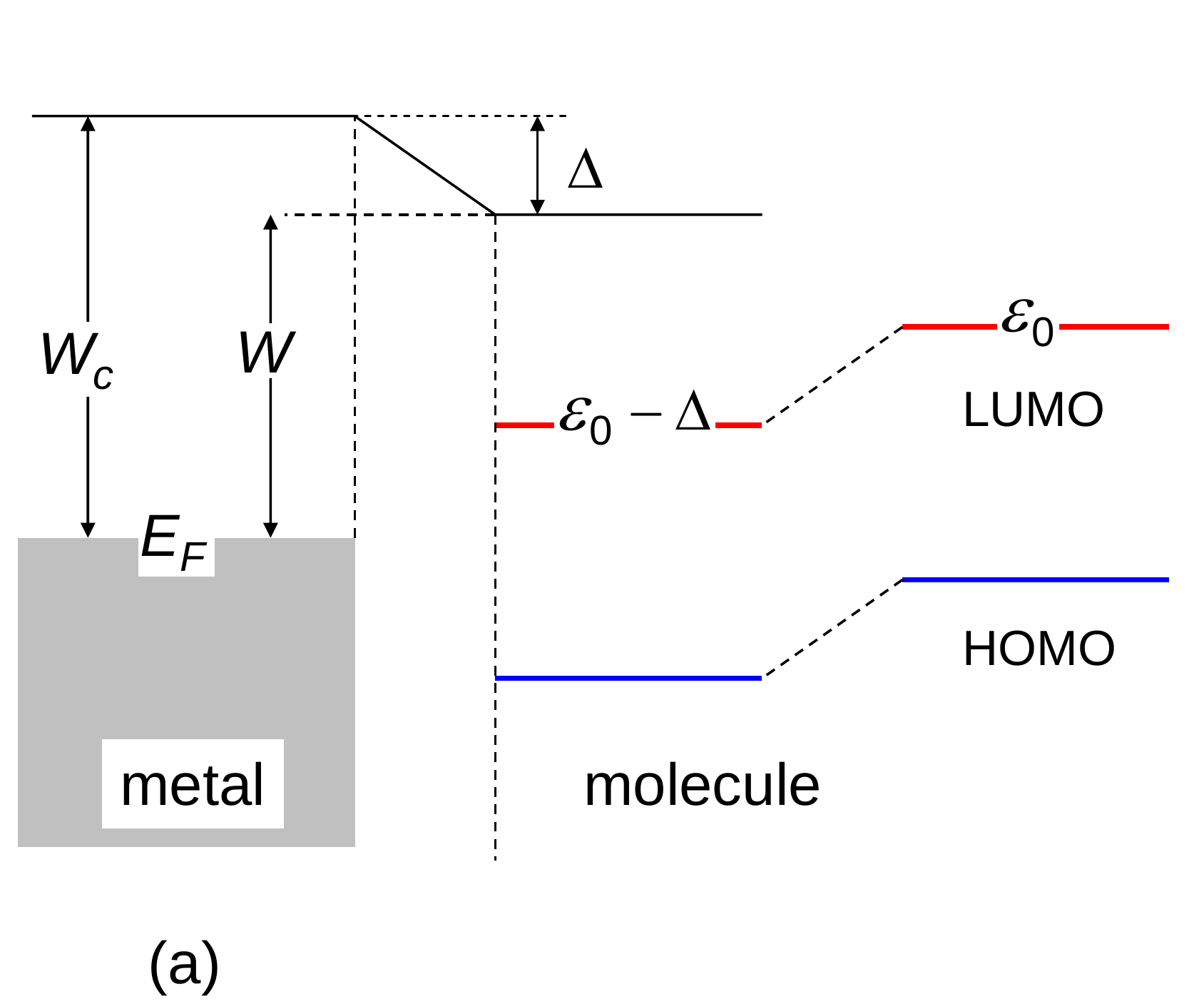}
\includegraphics[width=5.5cm,clip=true]{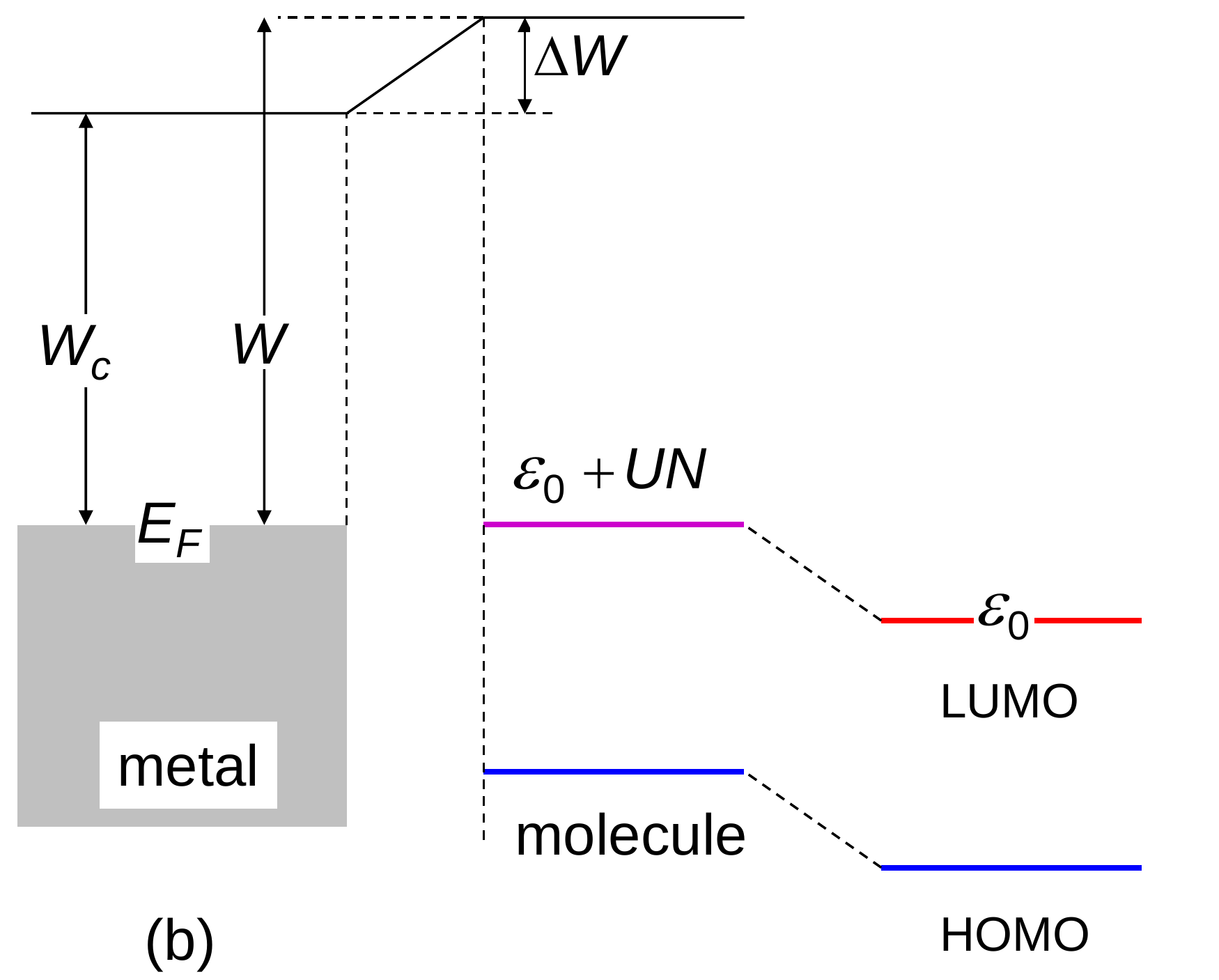}
\caption{(Color online) (a) The pillow effect results in a potential step $\Delta$, which lowers the molecular
levels close to the interface. (b) Charge transfer to the LUMO rises the molecular levels and pins
the Fermi level.} \label{diagram}
\end{figure}

If the molecular layer is in equilibrium with the metal substrate, one has $\epsilon_M(N) = E_F$,
with $E_F$ the Fermi level of the metal. This determines the occupation number
\begin{equation}
N = \frac{E_F - \epsilon_0}{U}. \label{eq:occnu}
\end{equation}
The idea is illustrated by Fig~\ref{diagram}(b). If $N \neq 0$, the molecular layer is charged and with the screening charge in the metal substrate this leads to a dipole layer. It results in a potential step at the
interface, which we parameterize as $NF$ with $F$ the potential step normalized per electron
transferred to a molecule. The work function then becomes
\begin{equation}
W = W_c + NF, \label{eq:wn}
\end{equation}
with $W_c = -E_F$ the work function of the clean metal surface.

So far we have not yet taken the pillow effect into account, which lowers the work function of the clean metal surface from $W_c$ to $W_c - \Delta$. As discussed in the previous section, the
pillow effect leads to an interface dipole that is localized mainly between the surface and the
molecule, see Fig.~\ref{delta n}(c). Therefore, it is consistent to apply the potential step to the molecular levels and replace $\epsilon_0$ by $\epsilon_0 - \Delta$ in Eq.~(\ref{eq:occnu}). The idea is illustrated by Fig~\ref{diagram}(a). Replacing $W_c$ with $W_c - \Delta$ in Eq.~(\ref{eq:wn}) and using Eq.~(\ref{eq:occnu}) then gives
\begin{equation}
W = (W_c- \Delta)(1-\frac{F}{U}) - \epsilon_0 \frac{F}{U}. \label{eq:W1}
\end{equation}
The $S$ parameter, Eq.~(\ref{S_parameter}), is then given by
\begin{equation}
S =  1-\frac{F}{U}. \label{eq:W1a}
\end{equation}

Since $0 \leq N \leq 2$, these expressions are valid if $\Delta - \epsilon_0 - 2U \leq W_c \leq
\Delta - \epsilon_0$. If the clean metal work function falls outside these bounds, one has
\begin{equation}
W = W_c - \Delta;\; S = 1, \label{eq:W2}
\end{equation}
for $W_c > \Delta - \epsilon_0$. The molecular level is then unoccupied, i.e. $N=0$ and only the
pillow effect is operative. If $W_c < \Delta - \epsilon_0 - 2U$, one has $N=2$. The molecular
level is fully occupied, which leads to
\begin{equation}
W = W_c - \Delta + 2F;\; S = 1. \label{eq:W3}
\end{equation}

From purely electrostatic considerations one has $0 < F \leq U$,\cite{Jackson:book} and $0 \leq S \leq 1$, cf. Eq.~(\ref{eq:W1a}). The expressions of Eqs.~(\ref{eq:W1}) and (\ref{eq:W1a}) can be simplified considerably if charge on the molecular layer and its counter charge in the metal substrate can be modeled by a plane capacitor. In that case we have $U = F = e^2/C$, where $C$ is the ``capacitance'' of the molecule,\cite{Sabin:ijqc00,Jackson:book} which leads to the simple expressions
\begin{equation}
W =  - \epsilon_0;\; S =  0. \label{eq:W4}
\end{equation}
In this limit one has perfect pinning, i.e. the work function is determined by the
molecular level only. It is independent of the metal and of the pillow effect.

This simple model can be used to qualitatively describe the work functions in
Figs.~\ref{results_LDAGGA} - \ref{results_perylene}. The simplest case is if $W_c > \Delta -
\epsilon_0$, i.e. if the work functions of all the metals considered are too high with respect
to the position of the LUMO level. $W$ and $S$ are then given by Eq.~(\ref{eq:W2}), i.e. the work
function is simply shifted with respect to the work function of the clean metal surface. Benzene
adsorbed on metal surfaces is such a case, see Fig.~\ref{results_benzene}. Since the work function
shift is determined by the pillow effect, one expects it to be sensitive the distance between the
molecule and the surface, which can be observed in Fig.~\ref{results_benzene}.

If $W_c \leq \Delta - \epsilon_0$, the LUMO reaches the Fermi level of the metal. $W$ and $S$ are
given by Eq.~(\ref{eq:W1}), and in the simple plane capacitor model by Eq.~(\ref{eq:W4}). A
close-packed monolayer of planar molecules, such as in the herringbone structure of PTCDA
(Fig.~\ref{structures}(b)), comes closest to a plane capacitor. Fig.~\ref{results_herringbone}
shows that indeed the work function is pinned at $\sim 4.7$ eV over a considerable range of metal
work functions and molecule-surface distances. If $W_c < \Delta - \epsilon_0 - 2U$, the LUMO
becomes fully occupied. $W$ and $S$ are given by Eq.~(\ref{eq:W3}) and the work function becomes
``unpinned''.

The energy at which this occurs depends upon the molecule-surface distance. Decreasing the
distance increases the pillow effect, i.e. it increases $\Delta$. Moreover it decreases $U$, since
at a shorter distance the screening by the metal substrate is larger. In the plane capacitor model
$U \propto 1/d$, where $d$ is the molecule-surface distance. The distance dependence of the
unpinning of the work function is observed in Fig.~\ref{results_herringbone}. At $d=3.6$ \AA\ the
work function is fully pinned, at $d=3.3$ \AA\ it becomes unpinned for Ca, and at $d=3.0$ \AA\ it
is unpinned for Ca and Mg.

The dilute structure for PTCDA has only $\sim \frac{1}{2}$ ML coverage (Fig.~\ref{structures}(a)).
This situation cannot be describe by a simple plane capacitor, and one has to
use Eq.~(\ref{eq:W1}). It gives a linear dependence of $W$ on $W_c$ with a slope $0<S<1$, which
can be observed in Fig.~\ref{results_LDAGGA}.

The behavior of perylene is consistent with the model given above. If $W_c > \Delta - \epsilon_0$,
only the pillow effect is operative and the LUMO is unoccupied, cf. Eq.~(\ref{eq:W2}). This holds
for Au and Ag in Figs.~\ref{results_perylene}(a) and (b). If $W_c \leq \Delta - \epsilon_0$, we
observe pinning. For a close-packed monolayer the plane capacitor model explains the pinning
(Eq.~(\ref{eq:W4})) observed for Mg and Ca in Fig.~\ref{results_perylene}(b). For the dilute
structure with $\sim \frac{1}{2}$ ML packing density, Eq.~(\ref{eq:W1}) can be used to describe
the behavior for adsorption on the low work function metals.

\section{Summary and conclusions}\label{conclusions}

We study the interface dipole formation at interfaces formed by a monolayers of PTCDA, benzene and
perylene molecules with the Au, Ag, Al, and Ca(111) and the Mg(0001) surfaces, using
first-principles DFT calculations. The interface dipoles are monitored by calculating the change
of the surface work function upon adsorption of the molecular layer. Molecular packing densities
corresponding to $\frac{1}{2}$ ML and 1 ML coverage are considered and the distance between the
molecules and the surfaces is varied to establish the dependence of the work function on these
parameters.

Adsorption of PTCDA in a densely pack structure leads to pinning ($S=0$) of the work function at
$\sim 4.7$ eV for a range of metal substrates and molecule-substrate distances, in good agreement
with experimental observations. The interface dipoles that are created upon adsorption compensate
for the differences between the work functions of the different clean metal surfaces. Along the
series Ag, Al, Mg, Ca the interface dipole generated by PTCDA adsorption increases, which is
consistent with an increasing transfer of electrons from the substrate to the PTCDA molecules.
The increased electron transfer also leads to a stronger bond between the molecule and the surface. Decreasing the packing density of the PTCDA molecules to $\frac{1}{2}$ ML decreases the pinning effect, but it still gives a linear dependence between the work function of the adsorbed layer and that of the clean metal surfaces ($S\approx 0.5$). Adsorption of PTCDA on Au(111) leads to a very weak bond and an interface dipole that has an opposite sign, as compared to the other surfaces. Here the pillow effect is dominant, which pushes the electrons into the metal substrate.

Adsorption of benzene results in a reduction of the work function, irrespective of the substrate,
in agreement with experiments. This reduction is in the range 0.2-0.8 eV, depending upon the
distance between the molecule and the surface. At a fixed distance $S\approx 0.9$. In the case of
benzene adsorption only the pillow effect is operative. The latter, as well as the effect of charge
transfer to the molecule are observed in adsorption of perylene. Adsorption of a full ML of
perylene molecules on low work function metals gives work function pinning ($S=0$) at $\sim 3.7$
eV. Adsorption on high work function metals gives the work function reduction characteristic of
the pillow effect with $S\approx 0.9$. The transition between the two regimes takes place for
substrate work functions in the range Mg-Al. Decreasing the packing density decreases the pinning
in the low work function regime.

A simple model inspired by Slater's transition state approach allows us to describe the
changes in the work function upon adsorption qualitatively. The model incorporates the charge transfer between the substrate and the molecular levels, the charging energy of the molecules, the pillow effect, and the interface dipole layer. It shows that for planar molecules the work function is pinned at a level that is determined by the molecules and not by the substrate. That level does not correspond to the molecular EA, but lies within the transport gap of the molecular material.

\begin{acknowledgments}
This work is part of the research programs of the Stichting voor Fundamenteel Onderzoek der Materie (FOM), financially supported by the Nederlandse Organisatie voor Wetenschappelijk
Onderzoek (NWO), and of NanoNed, a nanotechnology program of the Dutch Ministry of Economic Affairs. The use of supercomputer facilities was sponsored by the Stichting Nationale Computer Faciliteiten (NCF), financially supported by NWO.
\end{acknowledgments}


\begin{thebibliography}{137}
\bibitem[{fn:({\natexlab{a}})}]{fn:charging}
\bibinfo{note}{Note that the total energy of a molecular ion with $N$
  additional electrons, can be written as $E_\mathrm{tot}(N) =
  E_\mathrm{tot}(0) + \epsilon_0N + \frac{1}{2}U N^2$.}
\bibitem[{fn:({\natexlab{b}})}]{fn:pol}
\bibinfo{note}{The factor 2 originates from writing the total energy of a
  molecular ion with $N$ additional electrons, embedded in a crystal, as
  $E_\mathrm{tot}(N) = E_\mathrm{tot}(0) + \epsilon_0N +
  \frac{1}{2}U_\mathrm{bare} N^2 - P^{N-}$. If the polarization energy $P^{N-}$
  is determined purely by electrostatics, then $P^{N-}=N^2P^{-}$. Writing the
  total energy as in Ref.~\onlinecite{fn:charging} then leads to
  Eq.~(\ref{eq:ueff})}.
\bibitem[{fn:({\natexlab{c}})}]{fn:ICT}
\bibinfo{note}{In the integer charge transfer (ICT) model, $N$ is restricted to
  an integer.\cite{Fahlman:jpcm07,Braun:advmat09} In that case one needs to
  invoke a mixture of charged and uncharged molecules to interpret work
  function pinning,\cite{Davids:jap95} which may lead to a non-uniformly
  charged organic layer. The ICT model is developed for organic layers that are
  separated from the metal by a tunnel barrier. In the cases dicussed in this
  paper, the organic layer is in close contact with the metal and partial
  occupations can be achieved by hybridization of the molecular and the metal
  states.}
\end{thebibliography}
\end{document}